\documentclass[aps,tighten,amssymb]{revtex4}
\usepackage{graphicx}

\newcommand{\ltsim}{\;\raisebox{.4ex}{$<$}\hspace{-0.8em}\raisebox{-0.6ex}
                     {$\sim$}\;}
\begin{document}


\title{Electroproduction, photoproduction, and inverse electroproduction
    of pions in the first resonance region}

\author{Yurii S. Surovtsev$^a$%
\footnote{E-mail address: surovcev@thsun1.jinr.ru},
Tatiana D. Blokhintseva$^a$%
\footnote{E-mail address: blokhin@main1.jinr.ru},
Petr Byd\v{z}ovsk\'y$^b$%
\footnote{E-mail address: bydz@ujf.cas.cz},
Miroslav Nagy$^c$%
\footnote{E-mail address: fyzinami@savba.sk}\,}

\address{$^a$Joint Institute for Nuclear Research, Dubna 141980, Russia\\
$^b$Nuclear Physics Institute, Czech Academy of Sciences, 25068
\v{R}e\v{z}, Czech Republic\\
$^c$Institute of Physics, Slovak Academy of Sciences, D\'ubravsk\'a cesta
9, 845 11 Bratislava, Slovakia}


\begin{abstract}
Methods are set forth for determining the hadron electromagnetic
structure in the sub-$N\overline{N}$-threshold timelike region of the
virtual-photon ``mass'' and for investigating the nucleon weak
structure in the spacelike region from experimental data on the
process $\pi N\to e^+e^- N$ at low energies. These methods are
formulated using the unified description of photoproduction, 
electroproduction, and inverse electroproduction of pions in the 
first resonance region in the framework of the dispersion-relation 
model and on the basis of the model-independent properties of 
inverse electroproduction. Applications of these methods are 
also shown.
\end{abstract}

\pacs{PACS number(s): 11.30.Hv, 11.55.Fv, 11.80.Et}
\maketitle

\section{Introduction}
Processes of meson electroproduction have played an important role 
in studying the structure and properties of matter (see, {\it e.g.} 
review~\cite{Amaldi}). In the past few years, however, reactions
with production of dileptons in hadron-hadron and hadron-nucleus
collisions have drawn much attention \cite{ST-92,BSN-99,Titov,Berger,
Lutz,Bratk}.
In these reactions virtual photons, which materialize as dileptons
({\it e.g.}, the $e^+e^-$ pair), carry unique information on properties
of matter because the processes in which the particle structure is
formed proceed in the timelike region of the ``mass'' ($\lambda^2$) of
the virtual photon. Therefore, further investigation of these reactions
is necessary and promising in acquisition of new and perhaps unexpected
information about the form factors of hadrons and nuclei.

The inverse pion electroproduction (IPE), $\pi N\to e^+e^- N$, (or
dileptonproduction), being for a long time the only source of information
on the nucleon electromagnetic structure in the timelike region, has
been investigated both theoretically \cite{Geffen,BST-yaf75,Baldin,
ST-yaf72,Kulish,Furlan,Dombey,Tkeb,Smirn-Shum,Bietti,ST-yaf75} and
experimentally \cite{Samios,Berezh,Baturin} since the beginning of the
1960s. In Refs.~\cite{BST-yaf75,ST-yaf72,ST-yaf75}, we worked
out the method of extracting the pion and nucleon electromagnetic form
factors from IPE at low energies. This method has been successfully
applied in the analysis of experimental data on the nucleon and $^{12}$C
and $^7$Li nuclei ~\cite{Berezh,Baturin} and values of the form factors
were obtained for the first time in the timelike region of $\lambda^2$
ranging from 0.05 to 0.22 (GeV/c)$^2$. In Refs.~\cite{ST-92,Baldin},
the authors proposed to use IPE at intermediate (above $\pi N$ resonances)
energies and small $|t|$ to study the nucleon electromagnetic structure
and justified it up to $\lambda^2\approx m_\rho^2$.

Though experimental data \cite{Bardin} on the $p \overline p \to e^+ e^-$
process are now available, there still remains a wide range 
of $\lambda^2$ (up to $4m^2$), where the form factors cannot be measured 
directly in these experiments. On the other hand, the intense pion beams 
available now enable one to perform more detailed experiments on IPE 
aimed at both extracting the hadron structure and carrying out a multipole 
analysis similar to those for photoproduction and electroproduction
(see,{\it e.g.}, \cite{Dev-Lyth-Rank}). Such experiments can address 
interesting topics. For example, in the $\Delta(1232)(P_{33})$ region
it is challenging to verify the $\lambda^2$ dependence of the
color-magnetic-force contribution found in the constituent quark model
\cite{Goghilidze}. It is, therefore, worth recalling the earlier
discovered properties of the photoproduction, electroproduction, and IPE 
to consistently substantiate methods of studying the electromagnetic and 
weak structure of the nucleon on the basis of the IPE data in the first
resonance region, and to provide new results of this analysis.

Additional motive for studying IPE in the first resonance region is 
the possibility of investigating the nucleon weak structure by utilizing 
the same data as for the electromagnetic structure. This possibility is 
based on the current algebra (CA) description and on the remarkable 
property of IPE. In the IPE process the creation of the $e^+e^-$ pairs 
of maximal mass (at the ``quasithreshold'') is dominated by the Born 
mechanism, whereas the rescattering-effect contributions are at the 
level of radiative corrections up to the total $\pi N$ energy 
$w\approx 1.5$ GeV (the ``quasithreshold theorem'') \cite{ST-yaf72}. 
Due to this property, the threshold CA theorems for the pion
electroproduction and photoproduction can be justified in the case of 
IPE up to the indicated energy \cite{Kulish,Tkeb}. This allows one to 
avoid threshold difficulties when using the IPE data (unlike the 
electroproduction one) for extracting the weak form factors of the 
nucleon. Furthermore, in the case of IPE there is no strong kinematic 
restriction inherent to the $\mu$ capture and no kinematic suppression 
of contributions of the induced pseudoscalar nucleon
form factor to the cross sections of ``straight'' processes, such as
$\nu N\to lN$, present because of multiplying by the lepton masses.
Information on the pseudoscalar nucleon form factor $G_P$, which is
practically absent for the above reasons, is important because $G_P$
contains contributions of states with the pion quantum numbers and,
therefore, it is related to the chiral symmetry breaking. This would
enable us, {\it e.g.}, to test the Goldberger--Treiman relation.

Another aim of this paper is to draw attention of experimenters to the
process $\pi N\to e^+e^- N$ as a natural and unique laboratory for
investigating the hadron structure. 
One could use these processes for determining the baryon resonance
dynamics based on the study of the $\rho^0-\omega$ interference pattern
\cite{Titov,Lutz}. On the other hand, investigation of the exclusive 
reactions $\pi N\to l^+l^-N$ in experiments with
high-energy pions at large invariant mass of the dilepton and small
squared momentum transfer to the nucleon could provide access to
generalized parton distributions as suggested in \cite{Berger}.

This paper is organized as follows. In Sec.~II we give the basic 
formalism for the unified treatment of the reactions $\gamma+N\to\pi+N$,
$e+N\to e+\pi+N$, and $\pi+N\to e^++e^-+N$. In Sec.~III we present our
dispersion-relation model for unified description of these three reactions
and compare the calculated results with experimental data. To clearly 
explain our method, we choose a simple version of the model, which 
satisfactorily describes the data on the photoproduction and 
electroproduction. In Sec.~IV, we outline the method of determining the
nucleon electromagnetic form factors from low-energy IPE and discuss some
results of its application to analysis of the IPE data on the nucleon.
Section~V is devoted to extracting the pseudoscalar nucleon form factor
from the same IPE data, and interpretation of the results is given.
Concluding remarks are presented in Sec.~VI. Appendices present the 
relations between the amplitudes, derivation of the quasithreshold 
constraints for the multipole amplitudes on the basis of the principle 
of the first-class maximum analyticity, and explanation of the 
compensation effect. 

\section{Basic formalism}
We consider the reactions $\gamma+N\to\pi+N$, $e+N\to e+\pi+N$, and
$\pi+N\to e^++e^-+N$ in the framework of the unified model. This 
approach is natural because in the one-photon approximation, due to 
the $T$-invariance, research into these three processes is related to 
a study of the process ~$\gamma^* N\leftrightarrow\pi N$ with the hadron
current ~$J_\mu(s,t,\lambda^2)$, where~$\lambda^2=0$, ~$<0$, ~and ~$>0$~
correspond to the pion photoproduction, electroproduction, and IPE, 
respectively. This allows us to predict peculiarities of the IPE 
dynamics on the basis of a rich experimental material on the 
electroproduction and photoproduction and to test the reliability of 
the unified model for these three processes.

The $S$-matrix element for the electroproduction in the one-photon
approximation is
\begin{equation}
\label{Smatrix}
S_{fi}=\delta_{fi}+i(2\pi)^4\delta(p_1+k-p_2-q)\frac{em}
{\sqrt{2q_0p_{10}p_{20}k_{10}k_{20}}}\frac{m_e^2}{\lambda^2}
\varepsilon^\mu J_\mu(s,t,\lambda^2)\; ,
\end{equation}
where
\begin{equation}
\label{lept.current}
\varepsilon^\mu=\overline{u}(k_2)\gamma^\mu u(k_1)\; ,
\end{equation}
is the matrix element of the lepton electromagnetic current, $k_1=(k_{10},
{\vec k_1})$ ($k_2$), $p_1$ ($p_2$), and $q$ are the four-momenta of the
initial (final) electron, nucleon, and final pion, respectively. Momentum
of the virtual photon is $k=k_1-k_2$ ($k=k_1+k_2$ for IPE),
$k^2=k_0^2-{\vec k}^2\equiv\lambda^2$, and $s=(p_1+k)^2$ and $t=(k-q)^2$
are Mandelstam variables. Conservation of the lepton and hadron
electromagnetic currents implies $J_\mu k^\mu= \varepsilon_\mu k^\mu=0$.

Assuming the $T$-invariance, for the IPE process one must use the spinor
$v(k_1)$ instead of $u(k_1)$ in the lepton current (\ref{lept.current}).
Then $k$ is timelike, and ~$4m_e^2\leq\lambda^2\leq(\sqrt{s}-m)^2$~ is
the range of $\lambda^2$ values for fixed $s$.

The hadron current $J_\mu(s,t,\lambda^2)$ can be expanded using the six
independent covariant gauge-invariant structures $M_i$ \cite{BST-yaf75,Adler}:
\begin{equation}
\label{J:covar.expantion}
J_\mu\varepsilon^\mu=\sum_{i=1}^6 A_i(s,t,\lambda^2)
\overline{u}(p_2)M_iu(p_1),
\end{equation}
where
\begin{eqnarray}
\label{M_i}
\left.
\begin{array}{ll}
M_1=\frac{i}{2}\gamma_5\gamma^\mu\gamma^\nu F_{\mu\nu}\;,
&~~M_4=2i\gamma_5\gamma^\mu P^\nu F_{\mu\nu}-2mM_1\;,\nonumber \\
M_2=-2i\gamma_5 P^\mu q^\nu F_{\mu\nu}\;,
&~~M_5=-i\gamma_5 k^\mu q^\nu F_{\mu\nu}\;,\nonumber \\
M_3=i\gamma_5 \gamma^\mu q^\nu F_{\mu\nu}\;,
&~~M_6=i\gamma_5 k^\mu\gamma^\nu F_{\mu\nu}\;,\nonumber
\end{array}  \right.
\end{eqnarray}
with $F_{\mu\nu}=\varepsilon_\mu k_\nu-k_\mu\varepsilon_\nu$
and $P=\frac{1}{2}(p_1+p_2)$. The invariant amplitudes
$A_i(s,t,\lambda^2)~(i=1,\cdots,6)$ are free from kinematic constraints,
but $A_2$ and $A_5$ have a kinematic pole at $t=m_\pi^2+\lambda^2$. The
amplitudes $A_5$ and $A_6$ are absent in the photoproduction.

The matrix element (\ref{J:covar.expantion}) can be expressed through
the scalar c.m. amplitudes
\begin{equation}
\label{T-F}
\frac{m_e^2}{\lambda^2}\varepsilon^\mu J_\mu\ =
\chi^+_2F\chi_1\;,
\end{equation}
where $\chi_1$ and $\chi_2$ are the Pauli spinors, and
\begin{eqnarray}
\label{F:scal.expantion}
F=
&&i({\vec \sigma}\cdot{\vec \varepsilon} - {\vec \sigma}\cdot{\tilde{\vec k}}
{\tilde{\vec k}}\cdot{\vec \varepsilon})F_1 + {\vec \sigma}\cdot{\tilde{\vec
q}}\,{\vec \sigma}\cdot[{\tilde{\vec k}}\times{\vec\varepsilon}\,]F_2 +
i{\vec\sigma}\cdot{\tilde{\vec k}}({\tilde{\vec q}}\cdot{\vec \varepsilon} -
{\tilde{\vec q}}\cdot{\tilde{\vec k}}{\tilde{\vec k}}\cdot{\vec
\varepsilon})F_3+\nonumber\\
&&+i{\vec\sigma}\cdot{\tilde{\vec q}}\,({\tilde{\vec
q}}\cdot{\vec \varepsilon} - {\tilde{\vec q}}\cdot{\tilde{\vec
k}}{\tilde{\vec k}}\cdot{\vec \varepsilon}\,)F_4 +i{k^2\over k_0}{\vec
\sigma}\cdot{\tilde{\vec k}} {\tilde{\vec k}}\cdot{\vec \varepsilon}\,F_5
+i{k^2\over k_0}{\vec \sigma}\cdot{\tilde{\vec q}}\,{\tilde{\vec
k}}\cdot{\vec \varepsilon}\,F_6\;.
\end{eqnarray}
Here $\tilde{\vec q}$ and $\tilde{\vec k}$ are unit vectors. The
amplitudes $F_{1,2,3,4}$ describe the process with the transversal photons
and $F_{5,6}$ with the longitudinal ones. The relations between the
amplitudes $F_i$ and $A_i$ are listed in Appendix~A.

In the isotopic space $F_i$ (and $A_i$) are matrices
\begin{equation}
\label{F:iso-expantion}
F_i^{\alpha}=\tau_\alpha F_i^{(0)}+\delta_{\alpha3}F_i^{(+)}\pm\frac{1}{2}
[\tau_\alpha,\tau_3]F_i^{(-)}\; ,
\end{equation}
where the upper sign corresponds to $\gamma^*N\to\pi N$ and the lower one
to $\pi N\to\gamma^*N$ process.

For the physical processes, we have
\begin{eqnarray} \label{F:phys}
\left.
\begin{array}{llll}
F_i(\gamma^*p\leftrightarrow\pi^+n)&=\sqrt{2}(F_i^{(0)}+F_i^{(-)})\;,&
F_i(\gamma^*n\leftrightarrow\pi^-p)&=\sqrt{2}(F_i^{(0)}-F_i^{(-)})\;,
\\
F_i(\gamma^*p\leftrightarrow\pi^0p)&=F_i^{(0)}+F_i^{(+)},&
F_i(\gamma^*n\leftrightarrow\pi^0n)&=-F_i^{(0)}+F_i^{(+)}\;.
\end{array}
\right.
\end{eqnarray}

The differential cross section for the electroproduction is written down
in the following form taking into account a possibility of the
longitudinal polarization of the electron
\begin{eqnarray}
\label{cr.section:el-pr-n}
{d^3\sigma \over {d\Omega^L_e}{dk^L_{20}}{d\Omega_{\pi}}}&=&
{\alpha\over 2\pi^2}{k^L_{20}\over k^L_{10}}{{\bf k}_L\over (-\lambda^2)}{1\over
1-\epsilon}{{\bf q}\over {\bf k}}{1\over 4}[T_1 + \epsilon \cos 2\phi\;
T_2 +\nonumber\\ &&\sqrt{2\epsilon(1+\epsilon)}\cos \phi\;
\sqrt{-\lambda^2}\;
T_3-\epsilon \lambda^2\; T_4 -\xi\sqrt{2\epsilon (1-\epsilon)}\sin
\phi\;  \sqrt{-\lambda^2}\; T_5 ]\;,
\end{eqnarray}
where ~$\epsilon = [1-2({\bf k}^2_L/\lambda^2)\tan^2 (\theta_L/2)]^{-1}$,
${\bf k}\equiv\vert{\vec k}\vert$ and {\bf q} are magnitudes of the photon 
and pion c.m. momenta, respectively; $\phi$ is the azimuthal angle between
planes of the electron scattering and the reaction $\gamma^*N\to\pi N$;
${\bf k}_L$, $k_{10}^L$ ($k_{20}^L$) and $\theta_L$ are the photon-momentum
magnitude, the initial (final) energy and the electron scattering angle in 
lab frame, respectively; and $\Omega^L_e$ and $\Omega_{\pi}$ are the solid 
angles of the scattered electron in lab and pion in c.m. of the $\pi N$ 
system, respectively. Degree of the longitudinal polarization of 
the electron is $\xi={\vec l}\cdot{\vec k_1}/{\bf k_1}$ where ${\vec l}$ 
is the polarization vector of the electron in its rest system.

The IPE differential cross section reads
\begin{eqnarray}
\label{cr.section:IPE}
{d^3\sigma \over {d\Omega^{\gamma}{d\Omega^e}{dk^2}}}&=&
{\alpha\over16\pi^2}{1\over \lambda^2}{{\bf k}\over {\bf q}}{1\over 4}
[(1+\cos^2\theta)\; T_1 - \sin^2\theta \cos 2\phi\; T_2 +
\nonumber\\
&&\lambda\sin2\theta \cos \phi\; T_3 + \lambda^2\sin^2\theta\;  T_4 +
\xi \lambda\sin\theta \sin \phi\; T_5]\;,
\end{eqnarray}
where $\theta$ is the angle between the momenta of the final nucleon and
electron in the $e^+e^-$ c.m. system, $\phi$ is the angle between planes
of the reaction $\pi N\to \gamma^* N$ and the $e^+e^-$ pair production,
$\Omega^{\gamma}$ is the solid angle of the virtual photon production in
the c.m. of the $\pi N$ system, and $\Omega^e$ is the solid angle of
electron in the c.m. of the $e^+e^-$ pair.
$T_1$ in formulas (\ref{cr.section:el-pr-n}) and (\ref{cr.section:IPE})
describes the processes $\gamma^*N\leftrightarrow \pi N$ with the
unpolarized transversal photons, $T_2$ characterizes the asymmetry of
contributions of the transversal virtual photons linearly polarized in the
plane of the $\gamma^*N\leftrightarrow \pi N$ and normally to it, $T_4$ is
the contribution of the longitudinal photons, and  $T_3$ and $T_5$ are the
real and imaginary parts of the interference contribution of transversal
and longitudinal photon in the helicity basis. It is seen that $T_5$ is
related with the contribution of longitudinal polarization of electron to
the cross section.  The differential cross sections (\ref{cr.section:el-pr-n})
and (\ref{cr.section:IPE}), measured for the processes with electron
polarized in one direction, $\sigma(\xi=+1)$, and in the opposite one,
$\sigma(\xi=-1)$, generally should display the asymmetry $P_e$ 
given by the contribution of $T_5$ 
\begin{equation}
\label{P_e}
P_e=\frac{\sigma(+1)-\sigma(-1)}{\sigma(+1)+\sigma(-1)}\;.
\end{equation}

The quantities $T_i$ are related to the amplitudes $F_i$ in the c.m.
frame as
\begin{eqnarray}
\label{T_i-F_i}
&&T_1=\frac{\alpha m^2}{\pi s}\left[|F_1|^2+|F_2|^2- 2\cos\theta^\gamma
\mbox{Re}(F_1F_2^*)\right]+T_2,\nonumber\\
&&T_2=\frac{\alpha m^2}{2\pi s}\sin^2\theta^\gamma\left[|F_3|^2
+|F_4|^2-2\mbox{Re}(F_1F_4^*+F_2F_3^*+\cos\theta^\gamma F_3F_4^*)
\right],\nonumber\\
&&T_3=\frac{\alpha m^2}{\pi s}\sin\theta^\gamma\mbox{Re}\left[(F_2+
F_3+\cos\theta^\gamma F_4)F_5^*+(F_1+\cos\theta^\gamma F_3+
F_4)F_6^*\right],\\
&&T_4=\frac{\alpha m^2}{\pi s}\left[|F_5|^2+|F_6|^2+
2\cos\theta^\gamma\mbox{Re}(F_5F_6^*)\right],\nonumber\\
&&T_5=2\frac{\alpha m^2}{\pi s}\sin\theta^\gamma\mbox{Im}\left[(F_2^*
+F_3^*+\cos\theta^\gamma F_4^*)F_5+(F_1^*+\cos\theta^\gamma F_3^* +
F_4^*)F_6\right].\nonumber
\end{eqnarray}

\section{Dispersion-relation model}

To obtain reliable information on the nucleon structure, it is important
to find kinematic conditions under which the IPE dynamics is determined
mainly by a model-independent part of interactions, the Born one. To this
end, we utilize such general principles, as analyticity, unitarity, and
Lorentz invariance and the phenomenology of the processes $eN\to e\pi^{\pm}N$
and $\gamma N \leftrightarrow\pi^{\pm} N$, considered in the framework of
the unified (including IPE) model. We use a simple version of the model, 
which describes the experimental data satisfactorily with a minimal
number of parameters (electromagnetic form factors). This allows one to
carry out a simple analytic continuation from the spacelike to timelike
values of $\lambda^2$. For isovector amplitudes, the fixed-$t$ dispersion
relations without subtractions at finite energy are used, with the
spectral functions describing the magnetic excitation of the
$P_{33}(1232)$ resonance. The isoscalar amplitudes are described only by
the Born terms since the $P_{33}(1232)$ resonance does not contribute to
them due to the isospin conservation. In Refs.~\cite{BST-yaf75,Adler}, it
was shown that this model is successful in combined description of
experimental data on the pion electroproduction, photoproduction, and IPE 
in the total-energy region from the threshold up to 
$w=\sqrt{s}\approx 1.5$~GeV.
Moreover, this model is adequate for our purpose, namely to formulate the
method for determination of the form factors in the timelike region, for
the following reasons. First, in the dispersion-relation approach, the
spectral functions in the first resonance region are expressed through the
nucleon electromagnetic form factors and the phase shift of the
$\pi N$-scattering $\delta_{33}(w)$. This reduces considerably the number
of fitted parameters, which is especially important in the IPE analysis.
Second, in the $P_{33}(1232)$ region, the electric $E_{1+}$ and scalar 
$S_{1+}$ quadrupoles, which (as the magnetic dipole $M_{1+}$) describe
excitation of the resonance $P_{33}(1232)$, amount to not more than 15\%
of $M_{1+}$. For example, the photoproduction data \cite{Beck} give the
value $(-2.5\pm 0.1\pm 0.2)\%$ for the $E_{1+}/M_{1+}$ ratio.
Phenomenological results for the $\lambda^2$ dependence of the
$E_{1+}/M_{1+}$ and $S_{1+}/M_{1+}$ ratios are not stable yet and depend
upon the method of their extraction from electroproduction data (see
discussion of this point in Ref.~\cite{PDG-00}, p.~698). However, these
ratios do not exceed respectively, 7 and 15\% up to $\lambda^2=-4$
(GeV/c)$^2$. Based on the quark model calculations (see, {\it e.g.},
Refs.~\cite{Goghilidze,Capstick}), we suppose that an analogous situation
takes place also for $\lambda^2 > 0$, at least up to
$\lambda^2\approx $~0.3 (GeV/c)$^2$. Therefore, at the first stage of our
analysis we neglect the quadrupole excitations of $P_{33}(1232)$, which we
expect to be a good approximation for processes with unpolarised nucleons.

The conventional procedure of Reggeization provides us with behaviour
of the invariant amplitudes for $s\to\infty$ and small $|t|$ \cite{Sur}
$$A_i\sim s^{\alpha(t)-1}~(i\neq 5),~~~~~~~~~~~~~~~ A_5\sim s^{\alpha(t)}.$$
Consequently, in a complete $s$-channel description, we should write a
fixed-$t$ dispersion relation with one subtraction at finite energy for
the isovector amplitude $A_5^{(-)}$ and without subtractions for the
remaining amplitudes, taking into account their crossing properties.
However, the dispersion integrals with the spectral functions which
describe the magnetic excitation of the $P_{33}(1232)$ resonance  
converge very well already at $\approx$ 2~GeV for all the amplitudes $A_i$.
Therefore, we use the fixed-$t$ dispersion relations without subtraction
at the finite energy \cite{BST-yaf75,Adler} for all isovector amplitudes
\begin{eqnarray} \label{eq:DR} & &
A_i^{(\pm)}(s,t,\lambda^2)=\tilde{R}_5^{(-)}+c_5+R_i^{(\pm)}\Bigl(\frac{1}
{m^2-s}\pm\frac{\epsilon_i}{m^2-u}\Bigr)+\nonumber\\
& &~~~~~~+\frac{1}{\pi}\int_{(m+m_\pi)^2}^{\infty} ds^\prime~ \mbox{Im}
A_i^{(\pm)}(s^\prime,
t,\lambda^2)\Bigl(\frac{1}{s^\prime-s-i\varepsilon}\pm\frac{\epsilon_i}
{s^\prime-u}\Bigr),
\end{eqnarray}
and we take the Born approximation for the isoscalar ones
\begin{equation} \label{eq:A(0)}
A_i^{(0)}(s,t,\lambda^2)~=~R_i^{(0)}\Bigl(\frac{1}{m^2-s}+\frac{\epsilon_i}
{m^2-u}\Bigr),
\end{equation}
where~~$\epsilon_{1,2,4}=-\epsilon_{3,5,6}=1,\quad
u=2m^2+m_\pi^2+\lambda^2- s-t$,
\begin{eqnarray} \label{eq:Born} &&
R_1^{(\pm,0)}=-\frac{g}{2} F_1^{v,s} (\lambda^2),\qquad
~~~~~~~~R_2^{(\pm,0)}=\frac {gF_1^{v,s}
(\lambda^2)}{t-m_\pi^2-\lambda^2},\nonumber\\ &&
R_3^{(\pm,0)}=R_4^{(\pm,0)}=(-,+)\frac{g}{2} F_2^{v,s}
(\lambda^2), \qquad ~R_5^{(\pm,0)}=R_6^{(\pm,0)}=0,\\ &&
\tilde{R}_5^{(-)}=\frac{2g}{\lambda^2}~
\biggl[\frac{F_1^v(\lambda^2)}{t-m_\pi^2-\lambda^2}-\frac{F_\pi(\lambda^2)}
{t-m_\pi^2}\biggr]\;, \nonumber
\end{eqnarray}
and
\begin{equation} \label{eq:c5}
c_5=\frac{2}{m_\pi^2+\lambda^2-t}~\frac{1}{\pi}\int_{(m+m_\pi)^2}^{\infty}
\frac{ds^\prime}{s^\prime-m^2}~\lim_{t \to
{m_\pi^2+\lambda^2}}\bigl[(t-m_\pi^2- \lambda^2)\mbox{Im}
A_5^{(-)}(s^\prime,t,\lambda^2)\bigr]\;,
\end{equation}
with the $\pi N$ coupling constant $g^2/4\pi=14.6$ and the normalisation
of the form factors: $~~ F_1^{v,s}(0)=F_\pi(0)=1, ~~2m F_2^v(0)=3.7,~~ 2m
F_2^s(0)=-0.12$. The terms $\tilde{R}_5^{(-)}$ and $c_5$ belong only to
the amplitude $A_5^{(-)}$.

Note that $A_2$ and $A_5$ have the kinematic pole at $t=m_\pi^2+\lambda^2$.
However, these singularities are cancelled out kinematically because these
amplitudes enter into the matrix element through the combination
~$(s-m^2)A_2+\lambda^2 A_5$ which is equal to ~$2B_3-B_2$, where $B_2$ and
$B_3$ are the Ball amplitudes which have been proved to have no kinematic
singularities \cite{Adler}. In specific model calculations, the condition
\begin{equation} \label{eq:Adler-cond}
\lim_{t\to {m_\pi^2+\lambda^2}}(t-m_\pi^2-\lambda^2)\bigl[(s-m^2)A_2+
\lambda^2 A_5\bigr]=0\; ,
\end{equation}
which is ensured by the form of the term $c_5$ (\ref{eq:c5}), guarantees
absence of the singularity at $t=m_\pi^2+\lambda^2$.

The spectral functions $\mbox{Im} A_i^{(\pm)}(s^\prime,t,\lambda^2)$
are supposed to describe the magnetic excitation of the $P_{33}(1232)$
resonance
\begin{equation}
\label{eq:ImA} \mbox{Im}
A_i^{(\pm)}(s,t,\lambda^2)=\frac{4\pi}{3}{ 2 \choose {-1} }\frac
{mG_M^v(\lambda^2)\sin^2 \delta_{33}(w)}{wg{\bf q}^3
[(w+m)^2-\lambda^2]}~ a_i(w,t,\lambda^2),
\end{equation}
where ~$G_M^v=F_1^v+2mF_2^v$,~ $\delta_{33}(w)$ is the corresponding
phase shift of the $\pi N$-scattering amplitude, for which we utilize the
prescription from Ref.~\cite{Aznauryan},
and
\begin{eqnarray} \label{eq:ai} &&
a_i(w,t,\lambda^2)=\alpha_i(w,t)-\lambda^2 \beta_i(w),\qquad
a_{2,5}(w,t,\lambda^2)=\frac{\alpha_{2,5}(w,t)-\lambda^2
\beta_{2,5}(w)} {t-m_\pi^2-\lambda^2},~~~~\\ &&~~~~~~~i=1,3,4,6\nonumber
\end{eqnarray}
with the coefficients $\alpha_i,\beta_i$
\begin{eqnarray}
\label{coef.in-ai} \left.
\begin{array}{ll}
\alpha_1=\frac{1}{2}(w+m)[(w+m)q_0-m_\pi^2+3t],
&~~\beta_1=\frac{1}{2}(w+m-q_0), \\
\alpha_2=\frac{3}{2}(w+m)(m_\pi^2-t),
&~~\beta_2=\frac{1}{2}(w+m)+q_0, \\
\alpha_3=-\frac{1}{2}(w+m)(w+m-q_0)-\frac{3}{4}(m_\pi^2-t),
&~~\beta_3=-\frac{3}{4}\;, \\
\alpha_4=(w+m)(w+m+\frac{1}{2}q_0)-\frac{3}{4}(m_\pi^2-t),
&~~\beta_4=\frac{3}{4}\;,\\
\alpha_5=2(s-m^2)(w+m+\frac{1}{2}q_0)-\frac{3}{2}(w-m)(m_\pi^2-t),
&~~ \beta_5=\frac{3}{2}(w-m),\\
\alpha_6=-\frac{1}{2}(w+m)q_0-\frac{1}{4}(m_\pi^2+3t),
&~~\beta_6=-\frac{3}{4}\;.
\end{array}  \right.
\end{eqnarray}
Furthermore, the results of the photoproduction multipole analyses
\cite{Dev-Lyth-Rank} allow us to set $E_{0+}^{(0)}=0$ above the
$P_{33}(1232)$ energy. Prescriptions for the pion and nucleon
electromagnetic form factors are taken from Refs.~\cite{pion,nucleon}.
The model described above is the first (simple) reliable version of the
more complex model for unified treatment of contemporary experimental 
data on the pion photoproduction, electroproduction, and IPE in the 
energy region which spans from the threshold up to the second $\pi N$ 
resonance.

%
%
\begin{figure}[htb]
\begin{center}
\rotatebox{-90}{\scalebox{0.47}{\includegraphics{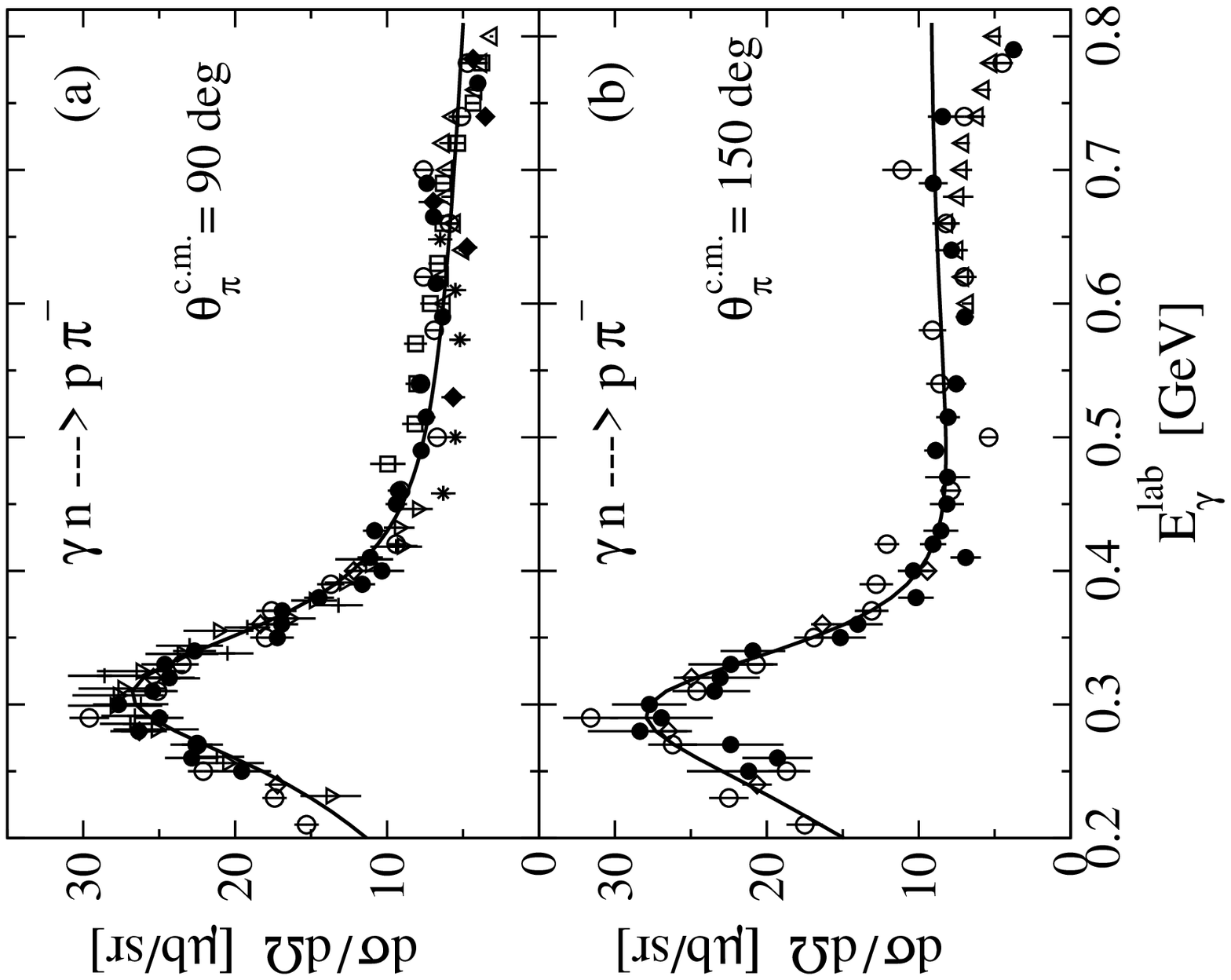}}}
\rotatebox{-90}{\scalebox{0.47}{\includegraphics{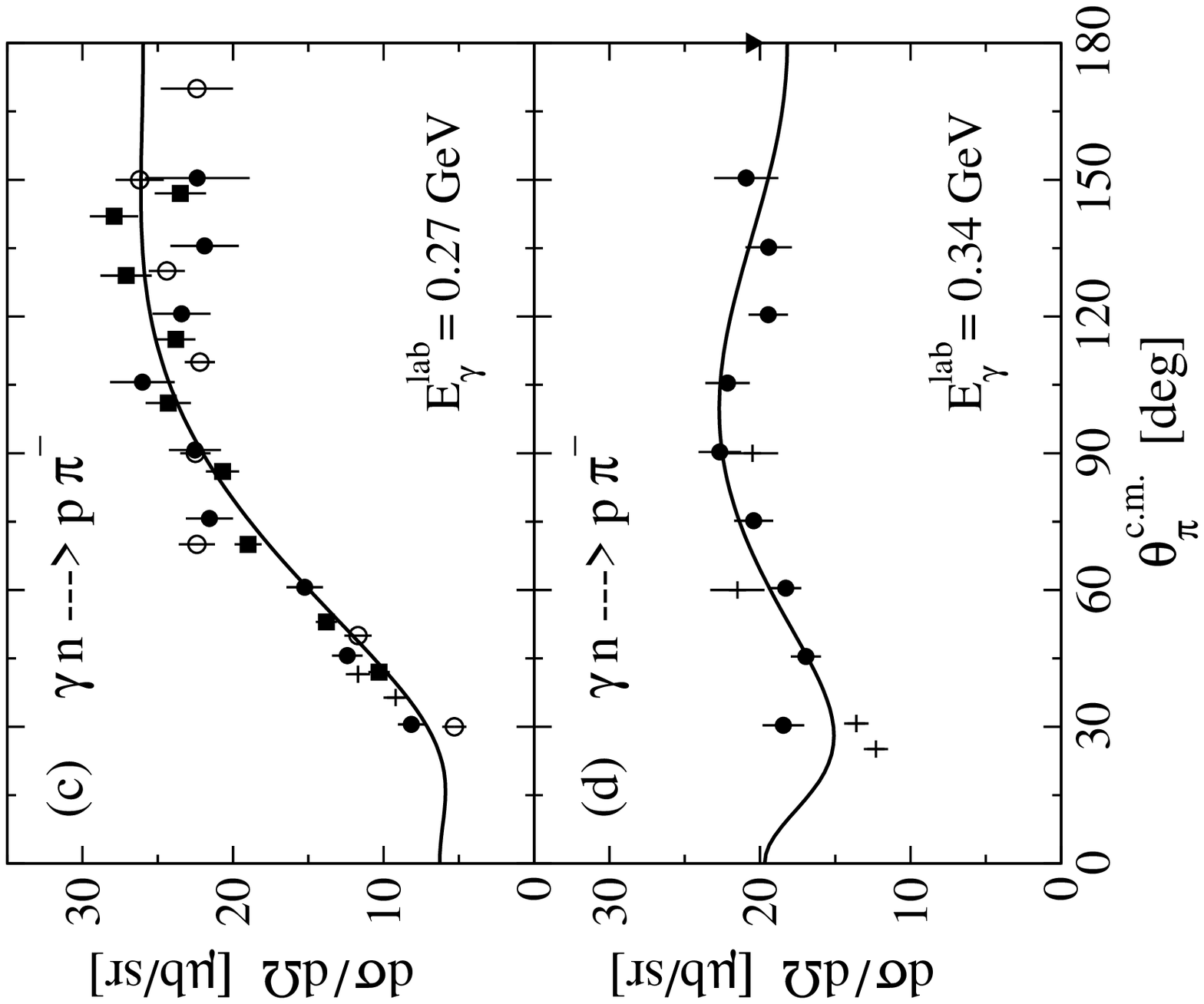}}}
\end{center}
\caption{Differential cross section for $\gamma n\to p\pi^-$
as a function of energy in (a) and (b) and pion angle in (c) and (d).
Experimental data are taken from Ref.~\protect\cite{DATA1}.}
\label{fig:ph-pr-n}
\end{figure}
%
%
\begin{figure}[htp]
\begin{center}
\rotatebox{-90}{\scalebox{0.47}{\includegraphics{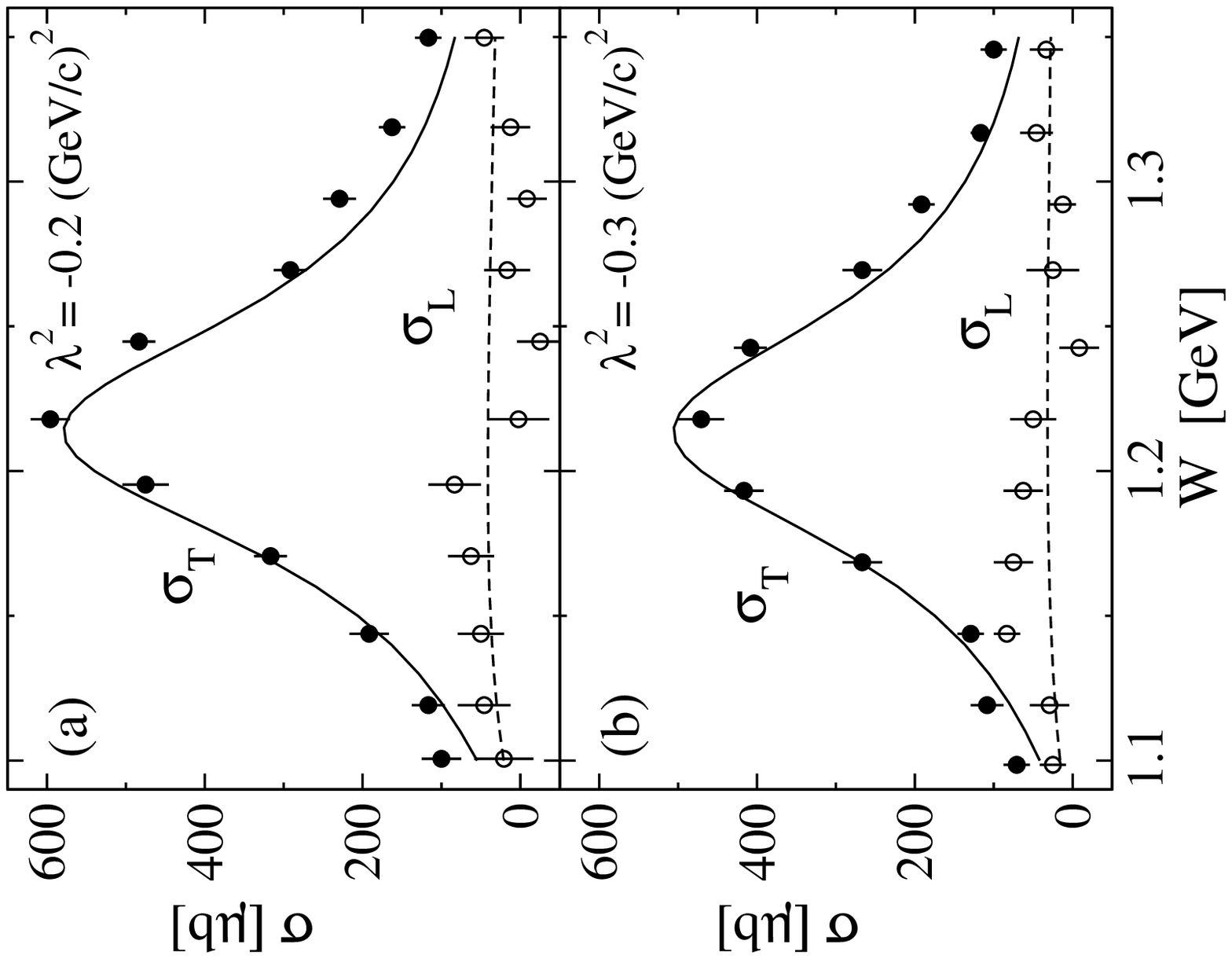}}}
\rotatebox{-90}{\scalebox{0.47}{\includegraphics{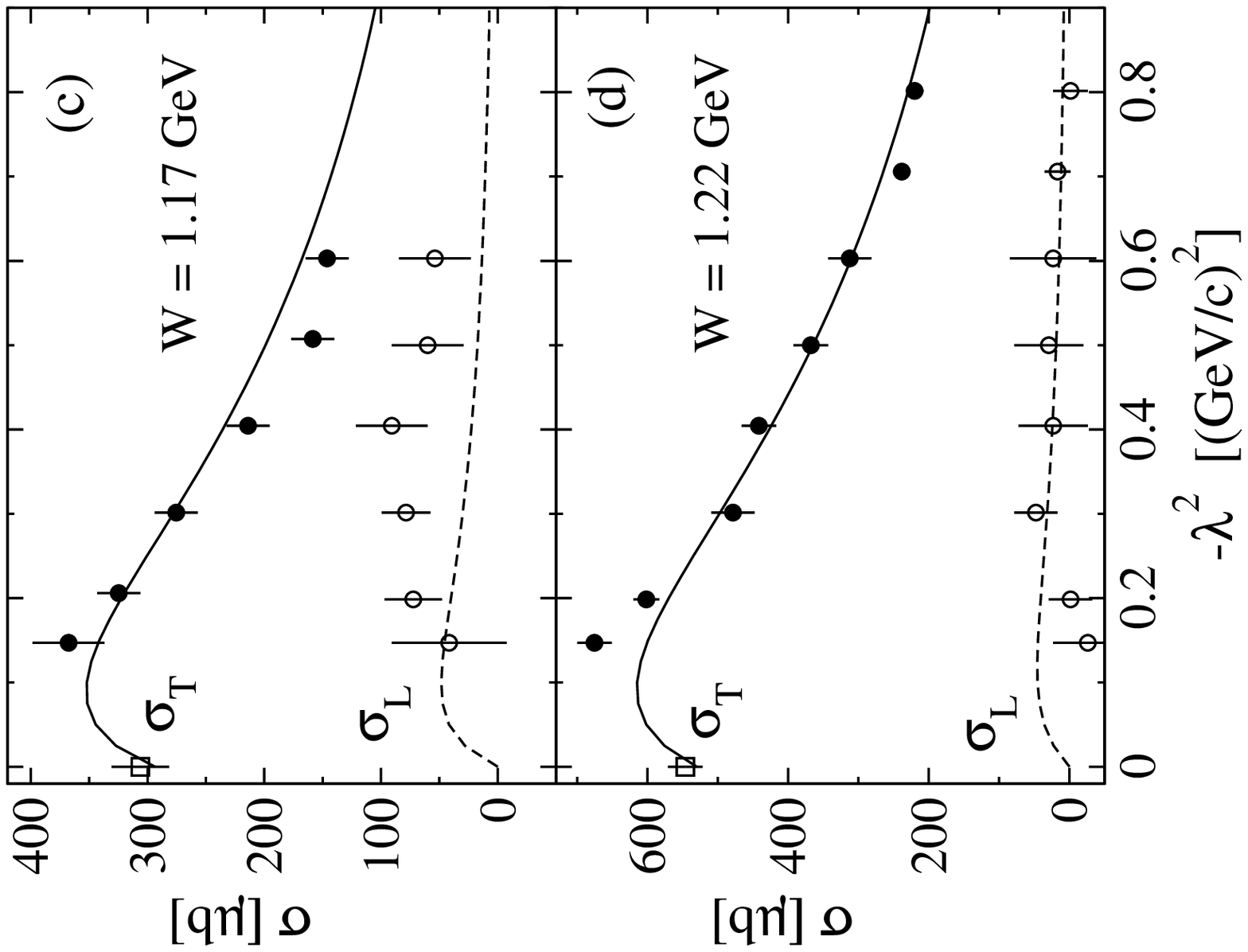}}}
\end{center}
\caption{$W$ and $\lambda^2$ dependences of the transversal $\sigma_T$ 
and longitudinal $\sigma_L$ parts of the cross section
(\protect\ref{d.cr.section:el-pr-n}) for the inclusive pion ($\pi^+$ and
$\pi^0$) electroproduction on the proton in the first resonance region.
Experimental data on the inclusive $p(e,e')$ process are from
Ref.~\protect\cite{Batzner} and the photoproduction points, squares in (c)
and (d), are from Ref.~\protect\cite{Fischer}.}
\label{fig:e-pr-n1}
\end{figure}
%
%
\begin{figure}[htp]
\begin{center}
\rotatebox{-90}{\scalebox{0.45}{\includegraphics{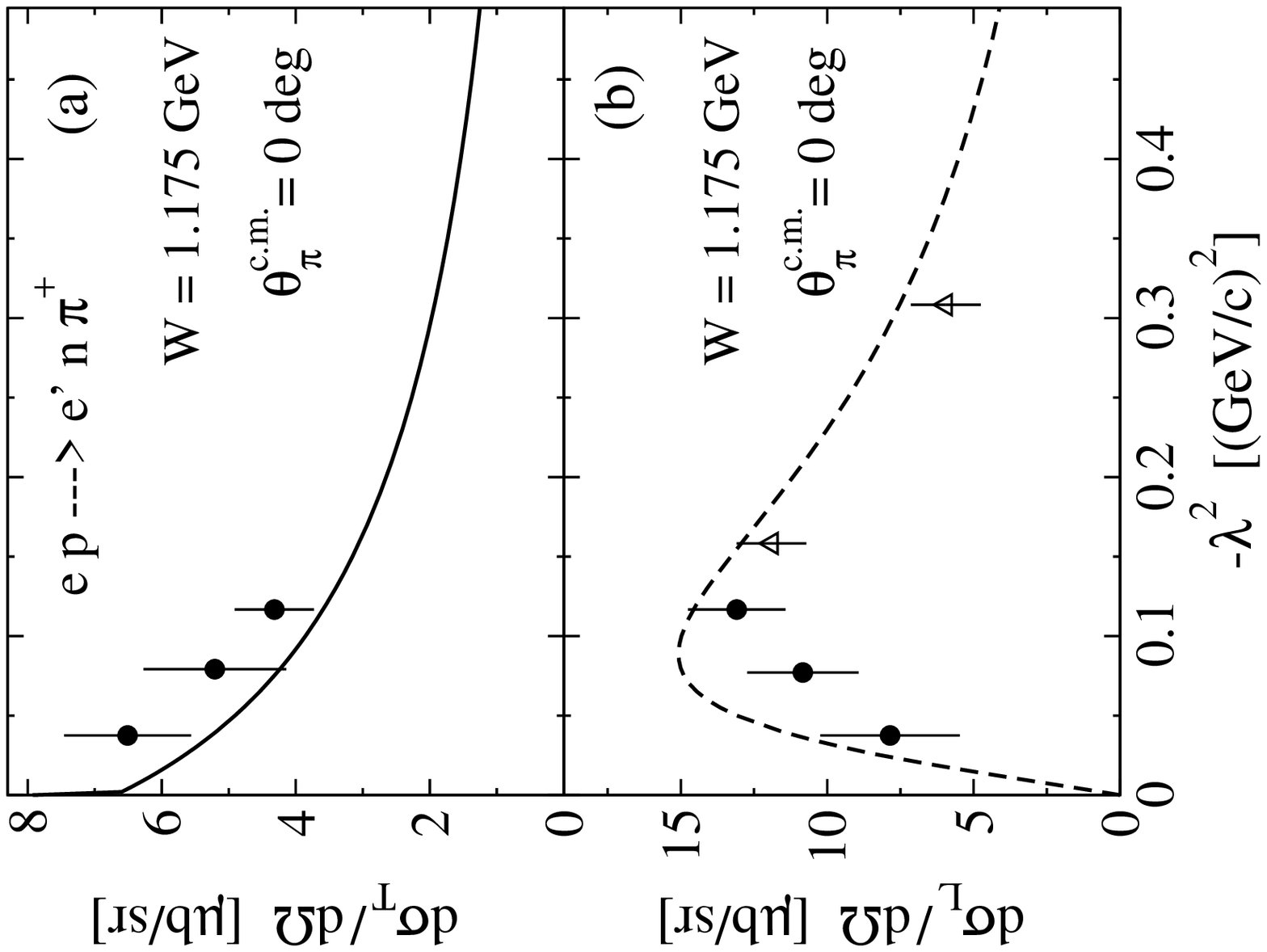}}}
\end{center}
\caption{$\lambda^2$ dependence of the differential cross section
for the forward electroproduction of pions by transversal (a) and
longitudinal (b) virtual photons. Experimental data are from
Refs.~\protect\cite{Bard77} and \protect\cite{Breuker}.}
\label{fig:e-pr-n2}
\end{figure}
In Fig.~\ref{fig:ph-pr-n} we compare results of our model for the
differential cross section with $\pi^-$-photoproduction data.
In Fig.~\ref{fig:e-pr-n1} we show comparison with the electroproduction
data on the proton for the transversal $\sigma_T$ and longitudinal 
$\sigma_L$ parts of the cross section
\begin{equation} \label{d.cr.section:el-pr-n}
{d^2\sigma \over {d\Omega^L_e\; dk^L_{20}}}=
{\alpha\over 2\pi^2}{{s-m^2}\over {2m(-\lambda^2)}}{k^L_{20}\over k^L_{10}}
{1\over 1-\epsilon}[\sigma_T(w,\lambda^2) + \epsilon\;
\sigma_L(w,\lambda^2)]\;.
\end{equation}
The differential cross sections for the forward electroproduction of pions
by transversal and longitudinal virtual photons are presented in
Fig.~\ref{fig:e-pr-n2}.

In general, we obtain quite a good agreement with the pion photoproduction 
and electroproduction data on unpolarized nucleons, especially in the region
of the $P_{33}(1232)$ resonance ($E_{\gamma}^{lab} \doteq 0.34$~GeV). In
the case of the $\pi^-$ photoproduction, the results are very good up to
$w\approx 1.5$~GeV ($E_{\gamma}^{lab} \approx 0.73$~GeV), see
Fig.~\ref{fig:ph-pr-n}.

It is obvious that our model can be further improved by including the
quadrupole excitation of the $P_{33}(1232)$ resonance ($E_{1+}^{\pm}$ and
$L_{1+}^{\pm}$) in the spectral functions. However, a still more elaborate
model should include, in addition to the quadrupole excitation,
contributions of other nucleon isobars and high-energy ``tails'' to the
absorption parts of the amplitudes to ensure a balanced consideration of
small corrections. Furthermore, analytic continuation of the corrected
absorption parts of the amplitudes into the unobservable region in the
dispersion integrals, $(m+m_\pi)^2\leq s\leq (m+\lambda)^2$ for
$\lambda^2>m_\pi^2$, requires use of the quasithreshold relations
(following from causality analyticity) between the electric and
longitudinal multipoles \cite{Sur} in which ``toroid'' multipoles appear
\cite{Dubovik}. On the contrary, the analytic continuation with the
approximation (\ref{eq:ImA}) is immediate. However, having in mind the
quality of contemporary experimental data, the above-stated simple model
seems to be quite sufficient (see
Figs.~\ref{fig:ph-pr-n}-\ref{fig:e-pr-n2}).

\section{Method of determining the nucleon electromagnetic 
structure from the low-energy IPE}

Application of the model to the calculations for IPE shows an interesting
growth of the relative contribution of the Born terms with $\lambda^2$ and
their dominance in the neighbourhood of the value
$\lambda^2=(\sqrt{s}-m)^2$ \cite{BST-yaf75}.
This approximate dominance of the Born terms has a model-independent
explanation. It is related to the quasithreshold theorem \cite{ST-yaf72}, 
which means that at the quasithreshold,
${\bf k}\to 0, ~\lambda^2\to\lambda_{max}^2=(\sqrt{s}-m)^2$, the IPE
amplitude becomes the Born one in the energy region from the threshold up
to $w\approx 1.5$~GeV. This remarkable dynamics of IPE distinguishes it
essentially from the photoproduction and electroproduction, where 
rescattering effects amount $\approx 40-50\%$.

Let us explain the quasithreshold behaviour of the IPE amplitude. As
${\bf k}\to 0$ the multipole amplitudes behave in the
following way
\begin{eqnarray} \label{q.thr.mult.ampl.}
&&M_{l\pm}\propto {\bf k}^l,~~~~~~~E_{l+}\propto{\bf k}^l,~~~~~~~L_{l+}\propto
{\bf k}^l,\nonumber\\
&& E_{l-}\propto {\bf k}^{l-2},~~~~~~~L_{l-}\propto {\bf k}^{l-2}.
\end{eqnarray}
Therefore, at ${\bf k}= 0$ only the electric ($E_{0+}$ and $E_{2-}$) and
longitudinal ($L_{0+}$ and $L_{2-}$) dipoles survive. In addition, the
number of independent dipole transitions diminishes to the two ones
at the quasithreshold due to the quasithreshold constraints
\begin{equation} \label{q-thr-constr}
E_{0+}=L_{0+}\;, \qquad~~~~~~E_{2-}=-L_{2-}\;,
\end{equation}
which arise from the causality (analyticity). The selection rules which
follow from the parity conservation and the value of the angular momentum
of the stopped virtual photon, $J=1$, restrict possible s-channel
resonances at the quasithreshold to the following sets:
$J^P={\frac{1}{2}}^-~[S_{11}(1535), S_{31}(1650),S_{11}(1700)$, {\it etc.}]
and $J^P={\frac{3}{2}}^-~[D_{13}(1520), D_{33}(1670)$, {\it etc.}].
Behaviour of the multipole amplitudes as ${\bf q}\to 0$ and ${\bf k}\to 0$
and the quasithreshold constraints among them are derived in Appendix~B.

Since the $s$- and $d$-wave $\pi N$ resonances are excited at energies
above 1.5~GeV, one can expect that the dipoles $E_{0+}$ and $E_{2-}$ are
dominated by the Born terms below this energy. This is in agreement with
the multipole analyses of charged pion photoproduction and confirmed by
the dispersion-relation calculations at $\lambda^2\neq 0$. We can,
therefore, conclude that in the quasithreshold region, the IPE amplitude is
given by the Born terms with accuracy better than $5\%$. At energies below
$\approx 1.5$~GeV, we can write for the quasithreshold IPE
\begin{eqnarray}
\label{q-thr-sigma}
\lim_{{\bf k}\to 0}~\frac{\bf q}{\bf k}~\frac{d^2\sigma}{d\lambda^2d\cos\theta} &
\approx & \frac{\alpha}{12\pi}~\frac{m^2}{(\sqrt{s}-m)^2}~\frac{1}{s}
~\bigl[(1+\cos^2 \theta)|E_{0+}^{Born}+E_{2-}^{Born}|^2+\nonumber\\
& &~~~~~~~~~~~~~~~~~~~~~~~~~~~~+\sin^2\theta|E_{0+}^{Born}-
2E_{2-}^{Born}|^2\bigr].
\end{eqnarray}

In the real experiment, however, one cannot realize strictly the
quasithreshold conditions and, therefore, the realistic model (presented
above) is needed. In choosing the optimal geometry of the experiment for
deriving the form factors, the ``compensation curves'' \cite{ST-yaf75} can
help. The curves are defined as curves in the $(s,t)$ plane along which
the differential cross section is given only by the Born terms. These
curves can be constructed by comparing photoproduction experimental data
with the Born cross section and employing the existence theorem for
implicit functions (more details on the compensation effect is given in
Appendix~C).

The method of determining the electromagnetic form factors from low-energy
IPE is, therefore, based on utilizing the quasithreshold theorem, the
realistic dispersion-relation model, and the compensation curve.
This method has been already used in the experiments on the nucleon and
nuclei $^{12}$C and $^7$Li \cite{Berezh,Baturin}.

\begin{table}[htb]
\caption{\label{tab1}Electromagnetic form factors obtained in
experiments on the nucleon. The virtual photon momentum $\lambda^2$ 
is given in units of the pion mass.}
\begin{tabular}{ccccccccccc} \hline
${\lambda^2}[m_{\pi}^2]$ & 2.77 & 2.98 & 3.44 & 3.75 & 4.00
& 4.47 & 4.52 & 5.28 & 5.75 & 6.11 \\ \hline
$F_1^v$ & 0.96 & 0.93 & 1.16 & 1.04 & 1.14
& 1.22 & 1.13 & 1.20 & 1.32 & 1.36 \\
$F^\pi$ & 0.91 & 0.85 & 1.04 & 0.91 & 0.99
& 1.04 & 0.95 & 1.01 & 1.12 & 1.16 \\
Error  & 0.10 & 0.09 & 0.10 & 0.08 & 0.16 & 0.10
& 0.09 & 0.09 & 0.10 & 0.08 \\
\hline
\end{tabular}
\end{table}
In Table~\ref{tab1}, we present values of the electromagnetic form factors
obtained in experiments on the nucleon. In Tab.~\ref{tab1}, the same
experimental errors are given for $F_1^v$ and $F^\pi$ because in this
$\lambda^2$-range these form factors can be connected with each other  via
the spectral function by the relation
$F_1^v(\lambda^2)-F^\pi(\lambda^2)=\bigtriangleup(\lambda^2)$.
The quantity $\bigtriangleup(\lambda^2)$, taken from the dispersion
calculations \cite{Hohler}, possesses a significantly smaller theoretical
uncertainty than the calculated quantities $F_1^v$ and $F^\pi$. This is
caused by cancellation of terms with the large uncertainties in the
spectral function of $\bigtriangleup(\lambda^2)$ which, therefore,
is dominated by the one-nucleon exchange contribution in the region
$4m_\pi^2\leq\lambda^2\ltsim 20m_\pi^2$. One can see that this result is
rather model-independent. Then the same experimental errors can be given
for $F_1^v$ and $F^\pi$. Of course, having high-statistics data for IPE
one need not use the relation of $F_1^v$ with $F^\pi$ in extracting these
quantities. The values of $F_1^v$ are quite consistent with the
calculations of the nucleon electromagnetic structure in framework of the
unitary and analytic vector-meson dominance model \cite{Dubnicka}.

Let us look at the possibility of investigating the form factor
$G^*_M(\lambda^2)$ of the $\gamma^*NN^*_{3/2}$ vertex at $\lambda^2 >0$.
Whereas a measurement of the differential cross section for the
electroproduction with unpolarised electrons in the $\Delta(1232)(P_{33})$
region at $\theta^{\gamma}$ out of the compensation curve allows one to
extract information about the form factors of the $\gamma^*NN^*_{3/2}$
vertex, this approach is not sufficiently effective for IPE because the
dominance of the Born mechanism which reaches $\approx 95\%$ at the
quasithreshold is also considerable at lower values of $\lambda^2$. It
turns out that the analysis of asymmetry $P_l$ (\ref{P_e}) in the
dilepton production near the quasithreshold gives us that chance.

We consider the quantity $T_5$, connected to the asymmetry $P_l$. It can
be expanded to series in ${\bf k}$ near the quasithreshold with taking into
account formulas (\ref{T_i-F_i}), (\ref{m-pole-expantion}) and
(\ref{gk-behavior})
\begin{equation} \label{T_5-expand}
T_5 = 2{\alpha m^2 \over \pi s}\sin {\theta^{\gamma}}\sum_{i=0}{\bf k}^it_i\;.
\end{equation}
The lowest term of the formal series in Eq.~(\ref{T_5-expand})
$$
t_0 = 3\cos \theta^{\gamma}[- {\rm Im} E^*_{2-}(L_{2-} + L_{0+})
+ {\rm Im} E^*_{0+}L_{2-}]_{{\bf k}=0}\; ,
$$
equals zero at the quasithreshold as it follows from the constraints
(\ref{q-thr-constr}). Behaviour of $T_5$ for ${\bf k}\to 0$ then is
\begin{equation} \label{T_5:theorem}
T_5\propto {\bf k}t_1\;,
\end{equation}
where
$$
t_1 = {\rm Im}~[({\tilde M}_{1+}-{\tilde M}_{1-})(E^*_{0+}+E^*_{2-})
+{\tilde E}_{1+}(E^*_{2-}-5E^*_{0+}) +{\tilde
L}_{1-}(E^*_{0+}-2E^*_{2-})- g{\tilde E}_{3-}E^*_{2-}]_{\bf k=0} .
$$
In the $P_{33}(1232)$ region, the amplitudes $E_{0+}$ and $E_{2-}$ are
dominated by the Born terms, and the imaginary parts of form factors are
negligible. Contributions of $E_{1+}$ and $L_{1-}$ amount to less than 15\%
of that of $M_{1+}$. Multipole amplitudes $M_{1-}$ and $L_{1-}$, related
to the excitation of the $P_{11}(1470)$ resonance, generally have to be
very small, as it is seen in analyses of photoproduction (especially on
the neutron). Therefore, with a good accuracy, we obtain
\begin{equation} \label{T_5}
T_5 \sim (E^{\rm Born}_{0+} + E^{\rm Born}_{2-}){\rm Im}~M_{1+}\; .
\end{equation}
Since the quasithreshold relations (\ref{gk-behavior}) seem to be
approximately realized in a rather wide interval in $\lambda^2$, the
asymmetry $P_l$ has to be sensitive to ${\rm Im}M_{1+}$ in the
$P_{33}(1232)$ region. The measurement of $P_l$ would, therefore, allow
one to study quantitatively the assumption about the dominance of the
magnetic dipole transition and to extract information on the form factor
$G^*_M(\lambda^2)$ at $\lambda^2 > 0$. This is possible because the
contribution of the background part of the amplitude to $T_5$ is
considerably suppressed in the quasithreshold region, the background part
reducing as ${\bf k}^2$ for ${\bf k}$ becoming zero.

\section{Method of determining the pseudoscalar form factor 
of nucleon from the quasithreshold IPE}

Now let us discuss another interesting possibility of investigating the
weak nucleon structure related to the nucleon Gamow--Teller transition
described by the matrix element
\begin{equation} \label{Gamow-Teller}
\left<N(p_2)|A_\mu^\alpha|N(p_1)\right>=\overline{u}(p_2)\frac{\tau^\alpha}{2}
\Bigl[\gamma_\mu G_A(\lambda^2)+k_\mu G_P(\lambda^2)\Bigr]\gamma_5u(p_1),
\end{equation}
where $A_\mu^\alpha$ is the axial-vector current, and $G_A(\lambda^2)$ and
$G_P(\lambda^2)$ are the axial and induced pseudoscalar form factors,
respectively.

An alternative description of IPE--which utilizes the current commutators,
PCAC, and completeness--allows one to derive a low-energy theorem at the
threshold, ${\vec q}=0$ and $\lambda^2\rightarrow m_\pi^2$, related to the
approximate chiral symmetry and $O(m_\pi^2)$ corrections. Minimization of
the continuum contribution at the quasithreshold justifies this approach
up to $w\approx 1.5$~GeV \cite{Furlan} with the continuum corrections
being practically the same as in the dispersion-relation description.
Then, one obtains for the longitudinal part of the $\pi^- p\to\gamma^* n$
amplitude at the quasithreshold retaining only the leading terms in
$\lambda^2/m^2, t/m^2$ \cite{Furlan}, such that
\begin{eqnarray} \label{long.part:CA}
E_{0+}-2E_{2-}&=&\frac{\lambda}{2m_\pi^2 f_\pi}
\sqrt{\frac{(w+m)^2-m_\pi^2}{mw}}\Bigl\{D(t)-
\Bigl(1+\frac{\lambda}{2m}\Bigr)D(m_\pi^2-\lambda^2) +\nonumber\\
&&+\frac{m_\pi^2}{2m}\Bigl[G_A(m_\pi^2-\lambda^2)-
\frac{t}{2m}G_P(m_\pi^2-\lambda^2)\Bigr]\Bigr\},
\end{eqnarray}
where the constant of the $\pi\to\mu+\nu_\mu$ decay $f_\pi$ is defined by
~$\left<0|A_\mu(0)|\pi(q)\right>=if_\pi q_\mu$, $D(t)=-2mG_A(t)+tG_P(t)$,
and the quasithreshold values of the variables are
$$w_{q.thr.}=m+\lambda, \quad t_{q.thr.}=(m_\pi^2-\lambda^2)\frac{m}
{m+\lambda}.$$

$G_A$ was measured in various experiments, first of all in
$\nu n\to {\mu}^-p$ and $\bar\nu p\to {\mu}^+n$. It is reasonable to use
first this result
\begin{equation} \label{G_A}
G_A(t)=G_A(0)\Bigl(1-t/M_A^2\Bigr)^{-2}, \qquad G_A(0)=-1.25,\quad
M_A=(0.96\pm 0.03) {\rm GeV}.
\end{equation}
However, it is difficult to obtain reliable information on $G_P$ in these
experiments since its contribution to cross sections is kinematically
suppressed (it is multiplied by the lepton masses). In the $\mu$-capture
and $\beta$-decay experiments, there is a kinematically restricted small
range of $|t|$, $\approx 0-0.01 ({\rm GeV/c})^2$, in which the weak form
factor can be determined, though with a large error. For example, its
value extracted from measurements of the $\mu$-capture in hydrogen
\cite{Bard} is ~~$G_P(-0.88m_\mu^2 )=-8.7\pm1.9$. $G_P$ has also been 
measured in the capture of polarized muons by $^{28}$Si nuclei
\cite{Brudanin}.

Equation~(\ref{long.part:CA}) shows that the kinematic suppression of the 
$G_P$ contribution is absent when the IPE data at the quasithreshold are 
used for extracting $G_P$. In this way, the pseudoscalar form factor 
$G_P(t)$ can be determined in the range up to $t\approx -15m_\pi^2$~ 
(which corresponds to $w\approx 1.5$~GeV). Adopting the quasithreshold
kinematics, one can avoid the threshold difficulties that are present when 
applying an analogous method to the analysis of electroproduction data.

%
%
\begin{figure}[htp]
\begin{center}
\rotatebox{-90}{\scalebox{0.37}{\includegraphics{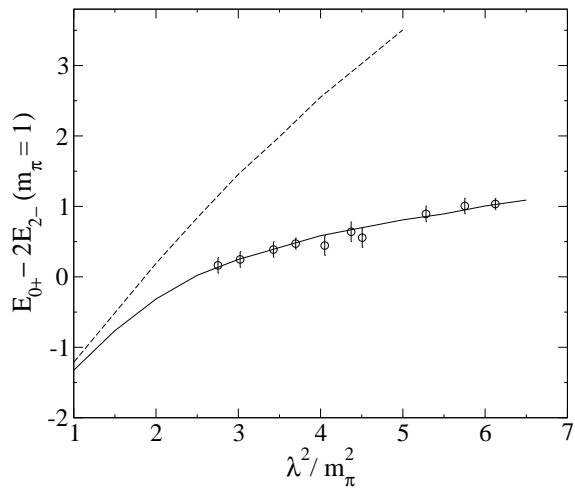}}}
\end{center}
\caption
{Comparison of CA calculations for $E_{0+}-2E_{2-}$ of the
$\pi^-p\to\gamma^* n$ process with experimental data: dashed and solid
curves correspond to the cases when $G_P$ is approximated by the pion
pole $G_P^\pi$ and the prescription (\protect\ref{G_P}) is assumed,
respectively. See the text for explanation of the data points.}
\label{fig:long-exp}
\end{figure}
Next we shall follow the method of Ref.~\cite{Tkeb}. First, using
the $F_1^v(\lambda^2)$ and $F_\pi(\lambda^2)$ values obtained in the
analysis of the IPE data on the nucleon \cite{Berezh}, we obtain 10 data
points for the longitudinal part of the $\pi^-p \to \gamma^* n$ amplitude
at the quasithreshold. These data points, which can be considered as
experimental ones, are depicted in Fig.~\ref{fig:long-exp}.

For $G_P(t)$ we chose the dispersion relation without subtractions:
\begin{equation} \label{G_P:DR}
G_P(t)=\frac{2f_\pi g_{\pi N}}{m_\pi^2-t}+\frac{1}{\pi}\int_{9m_\pi^2}^
{\infty} \frac{\rho(t^{\prime})}{t^{\prime}-t}dt^{\prime}.
\end{equation}
The residue at the pole $t=m_\pi^2$ is determined by the PCAC relation.
When only the $\pi$-pole term is considered in $G_P$, the result is
inconsistent with the experimental data as demonstrated in
Fig.~\ref{fig:long-exp} (dashed line). Therefore, the dispersion integral
in Eq.~(\ref{G_P:DR}) should be considered. It could be approximated by
the contributions of possible intermediate three-pion and resonance states
with the pion quantum numbers. However, since the contributions of
nonresonant three-particle states must be suppressed by the phase volume,
it is, therefore, reasonable to approximate the integral in (\ref{G_P:DR})
by a resonance-pole term. A satisfactory description is obtained if one
takes the following expression for $G_P(t)$ with the indicated values of
parameters:
\begin{equation} \label{G_P}
G_P(t)=G_P^\pi(t)-\frac{2f_{\pi^{\prime}} g_{\pi^{\prime} N}}
{m_{\pi^{\prime}}^2-t},\quad
2f_{\pi^{\prime}} g_{\pi^{\prime} N}=(1.97\pm 0.18)\;{\rm GeV},\quad
m_{\pi^{\prime}}=0.5 \;{\rm GeV},
\end{equation}
where ~$G_P^\pi(t)=2f_\pi g_{\pi N}/(m_\pi^2-t)$, the $\pi^{\prime}$
weak-decay constant $f_{\pi^{\prime}}$ is defined by
~$\left<0|A_\mu(0)|\pi^{\prime}(q^{\prime})\right>=
if_{\pi^{\prime}} q_\mu^{\prime}$, and $g_{\pi N}(=13.5)$ and
$g_{\pi^{\prime} N}$ are the coupling constants of the $\pi$ and
$\pi^{\prime}$ states with the nucleon, respectively. As seen from
the definitions of the weak-decay constants, one must expect that
~$f_{\pi^{\prime}}\ll f_\pi$ to reflect a tendency of another way (in
addition to the Goldstone one) in which the axial current is conserved 
for vanishing quark masses. This behaviour is demonstrated in various 
models with some nonlocality which describe chiral symmetry breaking
\cite{Volkov,Kalin}. Notice that the pole at $t=m_{\pi^\prime}^2$ in
Eq.~(\ref{G_P}), situated considerably lower than the poles of the known
contributing states $\pi^{\prime}(1300)$ and $\pi^{\prime}(1770)$, is
indispensable for describing the obtained experimental data on IPE.

%
%
\begin{figure}[htp]
\begin{center}
\rotatebox{-90}{\scalebox{0.42}{\includegraphics{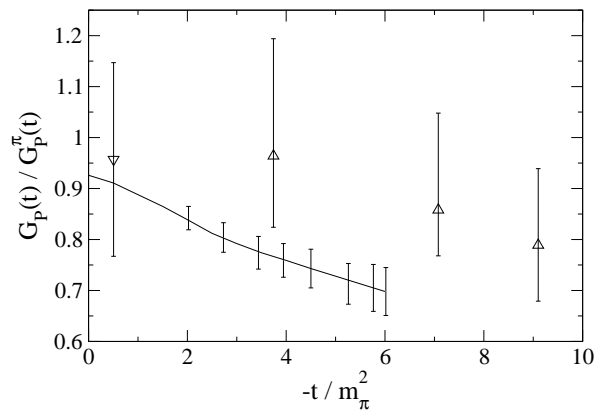}}}
\end{center}
\caption{Ratio $G_P(t)/G_P^\pi(t)$ as calculated with formula
(\protect\ref{G_P}). The points on the curve with the error bar indicate
the error corridor for this curve. Results of the data analysis on the
$\mu$ capture in hydrogen ($\bigtriangledown$) \protect\cite{Bard} and
on the $\pi^+$ electroproduction off the proton near the threshold
($\bigtriangleup$) \protect\cite{Choi} are also depicted.}
\label{fig:Gp-ratio}
\end{figure}
In Fig.~\ref{fig:Gp-ratio} we show the ratio $G_P(t)/G_P^\pi(t)$ in
comparison with the values obtained in $\mu$-capture on hydrogen
\cite{Bard} and in analysis of data on the $\pi^+$ electroproduction off
the proton near the threshold \cite{Choi}. The error bars on the solid
line indicate that the values of $G_P(t)$ determined by this method are 
of high accuracy. One can see that the results of Refs.~\cite{Bard,Choi} 
agree with the pion-pole dominance hypothesis in a large range of
momentum transfer unlike our result in which the hypothesis is valid only
in a narrow $t$ range, whereas outside the range the contribution of
continuum is considerable.

Note that the contributions of the pion radial excitations,
$\pi^{\prime}(1300)$ and $\pi^{\prime}(1770)$, which are rather distant
from this region, are suppressed and their consideration would only
slightly increase the mass of $\pi^{\prime}(500)$. The parameters of this
pole term in (\ref{G_P}) might be changed more considerably if one assumes
the channel $f_0(600)\pi$ with the recently discovered scalar $f_0(600)$
\cite{PDG-02}, due to the possible multichannel nature of this state. 
In any case, the conclusion will remain valid that the state with
$I^G(J^P)=1^-(0^-)$ in the range 500-800~MeV is needed for explaining the
obtained IPE data. Let us add that a possible signal of the charged state
of this isotriplet was observed in the $\pi^+\pi^-\pi^-$ system
\cite{Ivan} and interpreted as the first radial excitation of the pion in
the framework of the relativistic quark model \cite{Skachkov} based on the
covariant formalism for two-particle equations. Accepting this designation
for $\pi^\prime(500-800)$ and taking the estimate for the $\pi^{\prime}$
weak-decay constant in the Nambu -- Jona-Lasinio (NJL) model generalized 
by using effective quark interactions with a finite range,
$f_{\pi^{\prime}}=0.65$~MeV, we obtain $g_{\pi^{\prime} N}=1.51$. There
are no suitable theoretical calculations for this coupling constant now.
In the NJL model, the consideration of radial excitations of states
requires introducing some nonlocality. Since a successful calculation of
the $\pi N$ coupling constant in that model enforces one to go beyond the
framework of the tree approximation and take loop corrections into account
\cite{Nagy}, it seems that a satisfactory evaluation of
$g_{\pi^{\prime}N}$ in that approximation cannot avoid assuming some
nonlocality. Of course, a more reliable interpretation of $\pi^{\prime}$
requires investigation of other processes with $\pi^{\prime}$. Existence
of this state would also raise the question on its SU(3) partners.
A careful (re)analysis of the corresponding processes is, therefore,
desirable in this energy region.\vspace{-2mm}

\section{Conclusions}

We have demonstrated that a subsequent investigation of IPE is necessary
for extracting both unique information about the electromagnetic structure
of particles in the sub-$N\overline{N}$ threshold region of the timelike
values of $\lambda^2$ and the nucleon weak structure in the spacelike
region. The former is interesting especially now, {\it e.g.}, in
connection with discussion about the hidden strangeness of the nucleon
(see, {\it e.g.}, \cite{Gerasimov}) and quasinuclear bound $p\bar p$ state
\cite{Meshch}. Analyses of the experimental IPE data in the first $\pi N$
resonance region allow one to obtain the values of the form factor $F_1^v$
at timelike values of $\lambda^2$ which are quite consistent with the
calculations in the framework of the unitary analytic vector-meson
dominance model \cite{Dubnicka}. An inevitable step that is necessary for
the study of the electromagnetic structure of nucleon-isobar systems in
the timelike $\lambda^2$-region is a multipole analysis of IPE similar
to that for the photoproduction and electroproduction, 
{\it e.g.} \cite{Dev-Lyth-Rank}. At present, with the intense pion beams 
being available, it is possible to perform experiments aiming at carrying 
out that analysis. In a construction of the dispersion-quark model in 
the second and third $\pi N$ resonance region, the multichannel 
character of the nucleon isobars must be taken into account, {\it e.g.}, 
by the method of Ref.~\cite{Kamalov} or utilizing the proper uniforming 
variables \cite{KMS-nc}.

As we already mentioned, our method was used in the analysis of the
pion-induced dilepton production on the nucleon and light nuclei.
It is worth making some remarks about the analysis of the experiment
on the $^7$Li nucleus with a $\pi^+$ beam at 500~MeV/c \cite{Baturin}.
The missing mass analysis of the data has shown that about half of the 
events are related to disintegration processes of the nucleus which are 
dominated by the reaction\\
\hspace*{4cm} ~~~~~~~~~~~~~$\pi^+ + \;^7\mbox{Li}\; \to e^+ e^- + p + \;
^6\mbox{Li}\;$.\\
On analyzing this process it was assumed that the pion-nucleus amplitude
is determined by the neutron-pole mechanism, and the nuclear part
(the vertex function of the $\;^7\mbox{Li}\;\to \;^6\mbox{Li}\; + n$)
was calculated in the nucleon cluster model \cite{Avakov}.
The remaining events belong to the process\\
\hspace*{4cm}~~~~~~~~~~~~~$\pi^+ + \;^7\mbox{Li}\; \to e^+ e^- + \;
^7\mbox{Be}\;$.\\
When all the events were analyzed (with and without disintegration of the
nucleus), the cross section on the nucleus was supposed to be additively
connected with the cross section on the individual nucleon and nuclear
effects were taken into account via screening. In both cases, our model
was used for describing IPE on the individual nucleon. The obtained values
for $F_1^v$ are again quite consistent with the calculations in
Ref.~\cite{Dubnicka}. In the case of the reaction without disintegration
of the nucleus, one observed there the electromagnetic form factor of the 
nucleus in the timelike $\lambda^2$ region for the first time. 
Unfortunately, in the indicated analysis, unique information
on the electromagnetic structure of the nucleus in the timelike region
was lost. Generally, it seems at present that there is no satisfactory
concept of the electromagnetic form factors of the nucleus in the
timelike region. A satisfactory description must take into account both
a constituent character of the nucleus (and the corresponding analytic
properties) and more subtle (than the screening) collective nuclear
effects.

Finally, it should be noticed that a more reliable interpretation of
the observed state $\pi^{\prime}(500-800)$ requires solving a number of
questions, both theoretical and experimental. In the pseudoscalar sector,
states of various nature are possible, except for $q\bar q$, the $gg$
and $ggg$ glueballs, $q\bar q g$ hybrids, and multiquark states. However,
all the models and lattice calculations give masses of those unusual
states considerably greater than 1~GeV. Therefore, the most probable
interpretation of $\pi^{\prime}(500-800)$ seems to be the first radial 
pion excitation.\vspace{-2mm}


\begin{acknowledgments}
The authors are grateful to S.B. Gerasimov, V.A. Meshcheryakov,
and G.B. Pontecorvo for useful discussions and interest in this work.
Yu.S. acknowledges also support provided by the Votruba-Blokhintsev
Program for Theoretical Physics of the Committee for Cooperation of the
Czech Republic with JINR, Dubna. P.B. thanks the Grant Agency of the
Czech Republic, Grant No. 202/05/2142, and the Institutional Research Plan
AVOZ10480505. M.N. acknowledges the Slovak Scientific Grant Agency,
Grant VEGA No. 2/3105/23.
\end{acknowledgments}

\appendix
%
%
\section{}

The invariant amplitudes ${A}_i(s,t,\lambda^2)$ relate to the scalar
c.m. amplitudes $F_i(W,\cos\theta^{\gamma},\lambda^2)$ as follows
\begin{eqnarray} \label{A:F}
{K}{A}_1 &=&
(s-m^2-\lambda^2)\left(\frac{F_1}{p_{10}-m}-\frac{p_{20}+m}{\bf qk}F_2\right)+
\frac{m}{\bf qk}[2q_0\lambda^2+\nonumber\\
&& (t-m_{\pi}^2-\lambda^2)k_0]\left(\frac{F_3}{p_{10}-m}+
\frac{p_{20}+m}{\bf qk}F_4\right)-2m\lambda^2\left(\frac{F_5}{p_{10}-m}+
\frac{p_{20}+m}{\bf qk}F_6\right),\nonumber\\
{K}{A}_2 &=&
\frac{2}{t-m_{\pi}^2-\lambda^2}\left\{\lambda^2\left(\frac{F_1}{p_{10}-m}-
\frac{p_{20}+m}{\bf qk}F_2\right)+\frac{1}{2\bf qk}[(t-m_{\pi}^2-\lambda^2)k_0+
\right.\nonumber\\
&& \left.2q_0\lambda^2]\left[(w-m)\frac{F_3}{p_{10}-m}-\frac{p_{20}+m}
{\bf qk}(w+m)F_4\right]-\right.\nonumber\\
&& \left.\lambda^2\left[(w-m)\frac{F_5}{p_{10}-m}-
\frac{p_{20}+m}{\bf qk}(w+m)F_6\right]\right\},\nonumber\\
{K}{A}_3 &=& {K}{A}_4
+2w\frac{\bf k}{\bf q}\left(\frac{F_3}{p_{10}-m}+
\frac{p_{20}+m}{\bf qk}F_4\right),\\
{K}{A}_4 &=&
(w-m)\frac{F_1}{p_{10}-m}+(w+m)\frac{p_{20}+m}{\bf qk}F_2+
\frac{1}{2\bf qk}[(t-m_{\pi}^2-\lambda^2)k_0+\nonumber\\
&& 2q_0\lambda^2]\left(\frac{F_3}{p_{10}-m}+\frac{p_{20}+m}{\bf qk}F_4\right)-
\lambda^2\left[\frac{F_5}{p_{10}-m}+\frac{p_{20}+m}{\bf qk}F_6\right],\nonumber\\
{K}{A}_5 &=&
\frac{2}{t-m_{\pi}^2-\lambda^2}\left\{(s-m^2)\left(\frac{-F_1}{p_{10}-m}+
\frac{p_{20}+m}{\bf qk}F_2\right)+\frac{1}{2\bf qk}[(t-m_{\pi}^2-\right.\nonumber\\
&& \left.\lambda^2)(p_{10}+w)+2q_0(s-m^2)]\left[(w-m)\frac{-F_3}{p_{10}-m}+
\frac{p_{20}+m}{\bf qk}(w+m)F_4\right]+\right.\nonumber\\
&& \left.(s-m^2)\left[(w-m)\frac{F_5}{p_{10}-m}-\frac{p_{20}+m}{\bf qk}(w+m)F_6
\right]\right\},\nonumber\\
{K}{A}_6 &=&
(w+m)\frac{F_1}{p_{10}-m}+(w-m)\frac{p_{20}+m}{\bf qk}F_2-
\frac{1}{2\bf qk}[(t-m_{\pi}^2-\lambda^2)(p_{10}+w)+\nonumber\\
&& 2q_0(s-m^2)]\left(\frac{F_3}{p_{10}-m}+\frac{p_{20}+m}{\bf qk}F_4\right)-
(s-m^2)\left(\frac{F_5}{p_{10}-m}+\frac{p_{20}+m}{\bf qk}F_6\right),\nonumber
\end{eqnarray}
where ${K}=\frac{2s}{m}\sqrt{(p_{10}+m)(p_{20}+m)}$.
The inverse relations reads 
\begin{eqnarray} \label{F:A}
F_1 &=&
\frac{\sqrt{((w+m)^2-\lambda^2)((w+m)^2-m_{\pi}^2)}}{4mw}[(w-m){A}_1+
    (w-m)^2{A}_4-\nonumber\\
&&  \frac{1}{2}(t-m_{\pi}^2-\lambda^2)({A}_3-{A}_4)-\lambda^2{A}_6],
    \nonumber\\
F_3
&=& \sqrt{((w-m)^2-\lambda^2)((w-m)^2-m_{\pi}^2)}\frac{(w+m)^2-m_{\pi}^2}{8ms}
    [(s-m^2){A}_2+\nonumber\\
&&  \lambda^2{A}_5+(w+m)({A}_3-{A}_4)],\\ F_5 &=&
\frac{1}{4mw}\sqrt{\frac{(w+m)^2-m_{\pi}^2}{(w+m)^2-\lambda^2}}
    \left\{\frac{(t-m_{\pi}^2-\lambda^2)}{2}[(3s+m^2-\lambda^2){A}_2+
    2w({A}_3-{A}_4)+\right.\nonumber\\
&&  \left.(s-m^2+\lambda^2){A}_5]+[(w+m)^2-\lambda^2][{A}_1+(w-m)({A}_4-
    {A}_6)]+\right.\nonumber\\
&&  \left.(s-m^2+m_{\pi}^2)[(w+m)({A}_3-{A}_4)+(s-m^2){A}_2+
    \lambda^2{A}_5]\right\}\,.\nonumber
\end{eqnarray}
Formulas for the remaining amplitudes can be obtained from
Eqs.~(\ref{F:A}) using the formal substitution $w\to -w$ and symmetry
properties of the amplitudes $F_i$ under the substitution
\begin{equation} \label{F:w-w} F_2(w)=-F_1(-w),~~~
F_4(w)=-F_3(-w),~~~F_6(w)=F_5(-w). \end{equation}

%
%
\section{}

Here we demonstrate a derivation \cite{Sur} of the behaviour of the
multipole amplitudes as ${\bf q}\to 0$ and ${\bf k}\to 0$ on the basis of
the first-class maximum analyticity principle \cite{Chew}. According to
this principle, the total amplitude possesses only singularities that are
related to the dynamic processes and whose positions depend on masses of
intermediate (and external) states which are involved in these processes.
Then at the known arrangement of poles (and, therefore, of the threshold
branch points) all the other singularities are determined by the
systematic consideration of formulas for discontinuities in all channels.

Notice that the limit ${\bf k}\to 0$ is fulfilled in two cases:
$p_{10}\to m$ and $p_{10}\to-m$. In the former case, the limit is fulfilled 
in IPE at the quasithreshold when the virtual photon has a maximal mass, 
whereas in the latter, the limit is fulfilled in the $k^2$-channel, 
$e^+e^-\to \pi N_2{\bar N_1}$, at the stopped antinucleon $({\bar p}_{10}=m)$.

The amplitudes $F_i$ can be decomposed in terms of the multipole
amplitudes: the magnetic $M_{l\pm}~(j=l)$, electric $E_{l\pm}~(j=l\pm 1)$,
and longitudinal $L_{l\pm}~(j=l\pm 1)$ (or scalar $S_{l\pm}$ but due to
the current conservation, $L_{l\pm}=(k_0/{\bf k})S_{l\pm}$):
\begin{eqnarray}\label{m-pole-expantion}
F_1&=&\sum_{l=0}^{\infty}[lM_{l+}+E_{l+}]P^\prime_{l+1}
(\cos{\theta^{\gamma}})+
\sum_{l=2}^{\infty}[(l+1)M_{l-}+E_{l-}]P^\prime_{l-1}
(\cos{\theta^{\gamma}}),\nonumber\\
F_2&=&\sum_{l=1}^{\infty}[(l+1)M_{l+}+lM_{l-}]P^\prime_l
(\cos{\theta^{\gamma}}),\nonumber\\
F_3&=&\sum_{l=1}^{\infty}[-M_{l+}+E_{l+}]P^{\prime\prime}_{l+1}
(\cos{\theta^{\gamma}})+
\sum_{l=3}^{\infty}[M_{l-}+E_{l-}]P^{\prime\prime}_{l-1}
(\cos{\theta^{\gamma}}),\\
F_4&=&\sum_{l=2}^{\infty}[M_{l+}-M_{l-}-E_{l+}-E_{l-}]P^{\prime\prime}_l
(\cos{\theta^{\gamma}}),\nonumber\\
k_0F_5&=&\sum_{l=0}^{\infty}L_{l+}P^{\prime}_{l+1}(\cos{\theta^{\gamma}})-
\sum_{l=2}^{\infty}L_{l-}P^{\prime}_{l-1}(\cos{\theta^{\gamma}}),\nonumber\\
k_0F_6&=&\sum_{l=1}^{\infty}[L_{l-}-L_{l+}]P^{\prime}_{l+1}
(\cos{\theta^{\gamma}}).\nonumber
\end{eqnarray}
Notice the unphysical multipoles: $M_{0+}=E_{0-}=L_{0-}=E_{1-} =0$. The
multipole amplitudes are functions of $w={\sqrt s}$ and $\lambda^2$ only.
The properties of the amplitudes $F_i(w,\lambda^2,\cos\theta^\gamma)$ with
respect to the substitution $w\to -w$ (\ref{F:w-w}) result in the
corresponding symmetry relations for the multipole amplitudes
\begin{eqnarray} \label{Mac-Dowell}
-(l+1)M_{l+}(-w)&=&E_{(l+1)-}(w)+(l+2)M_{(l+1)-}(w),\nonumber\\
(l+1)E_{l+}(-w)&=&lE_{(l+1)-}(w)-M_{(l+1)-}(w),\\
L_{l+}(-w)&=&-L_{(l+1)-}(w).\nonumber
\end{eqnarray}
In the following, it is convenient to introduce new functions
${\tilde P}_l^{(n)}$ instead of the Legendre polynomial which have
singularities (due to
$\cos{\theta}^{\gamma}=(t-m_{\pi}^2-\lambda^2-2q_0k_0)/2{\bf qk}$)
\begin{equation} \label{tildeP}
{\tilde P}_l=({\bf qk})^lP_l,~~~~~{\tilde P}^{(n)}_l=({\bf qk})^{l-n}P_l^{(n)} .
\end{equation}

Taking into account the relation between
$F_i(w,{\theta^{\gamma}},\lambda^2)$ and $A_i(s,t,\lambda^2)$, we can
conclude that the quantities
\begin{eqnarray}
\label{nksF}
&&[(p_{10}+m)(p_{20}+m)]^{-1/2}F_1,~~~
[(p_{10}-m)(p_{20}-m)]^{-1/2}F_2,~~~
\frac{[(p_{10}-m)(p_{20}-m)]^{-1/2}}{p_{20}+m}F_3,\nonumber\\
&&\frac{[(p_{10}+m)(p_{20}+m)]^{-1/2}}{p_{20}-m}F_4,~~~
\left(\frac{p_{10}+m}{p_{20}+m}\right)^{1/2}F_5,~~~
\left(\frac{p_{10}-m}{p_{20}-m}\right)^{1/2}F_6
\end{eqnarray}
possess no kinematic singularities.

To guarantee this property of the quantities (\ref{nksF}) the multipole
amplitudes in (\ref{m-pole-expantion}) must have the form
\begin{eqnarray} \label{gk-behavior}
M_{l+}&=&\left(\frac{p_{20}+m}{p_{10}+m}\right)^{1/2}({\bf qk})^l\tilde{M}_{l+},~~~
M_{l-}=[(p_{10}+m)(p_{20}+m)]^{-1/2}({\bf qk})^l\tilde{M}_{l-},\nonumber\\
E_{l+}&=&\left(\frac{p_{20}+m}{p_{10}+m}\right)^{1/2}({\bf qk})^l\tilde{E}_{l+},~~~
E_{l-}=\left(\frac{p_{10}+m}{p_{20}+m}\right)^{1/2}({\bf q})^l{\bf k}^{l-2}
\tilde{E}_{l-},\\
L_{l+}&=&k_0\left(\frac{p_{20}+m}{p_{10}+m}\right)^{1/2}({\bf qk})^l
\tilde{L}_{l+},~~~
L_{l-}=k_0\left(\frac{p_{10}+m}{p_{20}+m}\right)^{1/2}({\bf q})^l{\bf k}^{l-2}
\tilde{L}_{l-}.\nonumber
\end{eqnarray}
Furthermore, we must ensure that the expression
$$\sum_{l=0}^{\infty} [l\tilde{M}_{l+}+\tilde{E}_{l+}]{\tilde
P}^\prime_{l+1}\over p_{10}+m$$
in $F_1$ has no kinematic singularities at $p_{10}=-m$, {\it i.e.}, we
must demand the following constraint:
\begin{equation} \label{c:m+e+.-m}
l\tilde{M}_{l+}+\tilde{E}_{l+}\stackrel{p_{10}\to -m}{\longrightarrow}
(p_{10}+m)\mu_l.
\end{equation}
In the exceptional case $l=0$, ($M_{0+}\equiv 0$) we obtain
\begin{equation} \label{c:E0+,-m}
\tilde{E}_{0+}\stackrel{p_{10}\to -m}{\longrightarrow}(p_{10}+m)\mu_0,
\end{equation}
{\it i.e.},
\begin{equation} \label{E0+}
E_{0+}=[(p_{10}+m)(p_{20}+m)]^{1/2}\mu_0.
\end{equation}
It is clear that a question about the kinematic zeros in $F_i$ is left
open. Since we are interested mainly in the $\lambda^2$ dependence,
we give only the kinematic $\lambda^2$ behaviour of the amplitudes $F_i$
as $p_{10}\to m$ and $p_{10}\to -m$, not writing down explicitly the
coefficients that depend only on $w$ at ${\bf k}=0$.

From (\ref{m-pole-expantion}) and (\ref{tildeP})-(\ref{E0+}), we have\\
a) As $p_{10}\to m$
\begin{equation} \label{Fbehavior.m}
F_1\sim F_4\sim F_5\propto 1,~~~~ F_2\sim F_3\propto
(p_{10}-m)^{1/2},~~~~ F_6\propto (p_{10}-m)^{-1/2},
\end{equation}
b) As $p_{10}\to -m$
\begin{equation} \label{Fbehavior.-m}
F_1\sim F_4\sim F_5\propto (p_{10}+m)^{1/2},~~~~F_2\sim F_3\sim
F_6\propto 1,~~~~ F_5\propto (p_{10}+m)^{-1/2}.
\end{equation}
Taking into account formulas (\ref{A:F}), (\ref{Fbehavior.m}) and
(\ref{Fbehavior.-m}), and the fact that the amplitudes $A_i$ have no
kinematic singularities (except for the pole at $\lambda^2=t-m_\pi^2$ in
$A_2$ and $A_5$), we conclude that the following constraints must be
fulfilled to satisfy the maximal analyticity:\\
a) As $p_{10}\to m$
\begin{eqnarray}
&&F_1 +\cos\theta^\gamma F_4 - k_0F_5\propto (p_{10}-m),
\label{c:F1,F3,F5.m}\\
&&\cos\theta^\gamma F_4 - k_0 F_6\propto (p_{10}-m)^{1/2},
\label{c:F4,F6.m}
\end{eqnarray}
b) As $p_{10}\to -m$
\begin{eqnarray}
&&F_2 + \cos\theta^\gamma F_4 - k_0 F_6\propto (p_{10}+m),
\label{c:F2,F4,F6.-m}\\
&&\cos\theta^\gamma F_3 - k_0F_5\propto (p_{10}-m)^{1/2}
\label{c:F3,F5.-m}.
\end{eqnarray}
Constraints (\ref{c:F1,F3,F5.m})-(\ref{c:F3,F5.-m}) and the $\lambda^2$
behaviour of the amplitudes $F_i$, (\ref{Fbehavior.m}) and
(\ref{Fbehavior.-m}), ensure the absence of kinematic singularities in the
amplitude $F$, Eq.~(\ref{F:scal.expantion}), as $p_{10}\to\pm m$ and
$t={const}$.

Taking into account the constraints (\ref{c:F1,F3,F5.m})-(\ref{c:F3,F5.-m})
and Eq.~(\ref{gk-behavior}) and using the recurrent formulas for the
Legendre polynomials, we obtain kinematic relations between the multipole
amplitudes at the quasithreshold and for $p_{10}\to -m$. The constraint
(\ref{c:F4,F6.m}) gives as $p_{10}\to m$
\begin{equation} \label{c:El-,Ll-.m}
(l-1)\tilde{E}_{l-}+k_0\tilde{L}_{l-}=(p_{10}-m)\kappa_l~~~~
(l\geq 2)\;,
\end{equation}
and
\begin{equation} \label{L1-}
L_{1-}=k_0[(p_{10}-m)(p_{20}-m)]^{1/2}\kappa_1\;.
\end{equation}
The constraint (\ref{c:F1,F3,F5.m}) gives
\begin{equation} \label{c:El+,Ll+.m}
(l+1)\tilde{E}_{l+}-k_0\tilde{L}_{l+}=(p_{10}-m)\tau_l\;,
\end{equation}
and (\ref{c:F3,F5.-m}), as $p_{10}\to -m$, results in
\begin{equation} \label{c:El+,Ll+.-m}
(l+1)\tilde{E}_{l+}-k_0\tilde{L}_{l+}=(p_{10}+m)\nu_l~~~~ (l\geq 1)\;,
\end{equation}
or
$$l(l+1)\tilde{M}_{l+}+k_0\tilde{L}_{l+}=(p_{10}+m)[(l+1)\mu_l-\nu_l]\;.$$
The multipole amplitude $L_{0+}$, behaves as
\begin{equation} \label{L0+}
\tilde{L}_{0+}=k_0[(p_{10}+m)(p_{20}+m)]^{1/2}\nu_0\;.
\end{equation}
Finally, the constraint (\ref{c:F2,F4,F6.-m}) gives the following
relations between the multipole amplitudes as $p_{10}\to -m$
\begin{eqnarray}
\tilde{M}_{l-}-\frac{(l-1)\tilde{E}_{l-}+k_0\tilde{L}_{l-}}
{p_{10}-m}&=&(p_{10}+m)\rho_l\;,~~~~ (l\geq 2)\;,\label{c:Ml-,El-,Ll-.-m}\\
\tilde{M}_{1-}-k_0\kappa_1&=&(p_{10}+m)\rho_1\;.\label{c:M1-,L1-.-m}
\end{eqnarray}

So, we derived the constraints for the multipole amplitudes at the
quasithreshold and as $p_{10}\to -m$. The relations of the first type
(\ref{c:El-,Ll-.m})-(\ref{c:El+,Ll+.m}) were also derived in
Refs.~\cite{Adler,Dubovik,Bjorken-Wal,Dev-Lyth} by a different
method that shades their kinematic nature.
The relations for $p_{10}\to -m$, analogous to (\ref{c:m+e+.-m}),
(\ref{c:El+,Ll+.-m})-(\ref{c:M1-,L1-.-m}), were also obtained in
a different way in Ref.~\cite{Dev-Lyth}. Note that since kinematic
conditions as $p_{10}\to -m$ ($\lambda^2\to(w+m)^2$) are obtained from the
quasithreshold kinematic configuration $p_{10}\to m$ ($\lambda^2\to(w-m)^2$)
through the substitution $w\to-w$, constraints for the multipole amplitudes
as $p_{10}\to -m$ can be derived from the ones at the quasithreshold with
the help of the symmetry relations for multipole amplitudes (\ref{Mac-Dowell})
taking into account the complete $\lambda^2$ behaviour of the latter as
$p_{10}\to m$ (\ref{gk-behavior}), (\ref{c:El-,Ll-.m})-(\ref{c:El+,Ll+.m}).
The relations for multipole amplitudes as $p_{10}\to -m$ are, of course,
not fulfilled in the reaction under consideration. A region of
their applicability is the c.m. system of pion and nucleon
in the $\lambda^2$ channel $e^+e^-\to\pi N_2{\overline N}_1$ at the
stopped antinucleon ($\overline{p}_{10}\to m$).

The relations between multipole amplitudes at the quasithreshold
are very important for our considerations because the condition
$p_{10}\to m$ is realized (at $\lambda^2>m_\pi^2$) in the investigated
reaction $\pi N\to e^+e^-N$ unlike the electroproduction and photoproduction.
The introduced quantities $\kappa_l$ (\ref{c:El-,Ll-.m}) and $\tau_l$
(\ref{c:El+,Ll+.m}) are not formal coefficients of expansion in a series
but they can be related to ``toroid'' multipole amplitudes $T_{l\pm}$
\cite{Dubovik} that have a definite physical meaning. The relations
(\ref{c:El-,Ll-.m}) and (\ref{c:El+,Ll+.m}) must be used with necessity
when carrying out analytic continuations of spectral functions in the
dispersion relations for IPE into the unphysical region
$(m+m_{\pi})^2\leq s^{\prime}\leq (m+\lambda)^2$. Therefore, it seems
that application (from the beginning) of toroid multipoles $T_{l\pm}$
instead of $L_{l\pm}$ is rather relevant in the timelike region.

%
%
\section{}

Here we shall find a region of variables $s,t$, and $\lambda^2$ where
the cross sections of the processes of pion photoproduction, electroproduction,
and IPE are described only by the Born terms. Let us write the
differential cross section of virtual photoproduction as a sum of two
terms
\begin{equation} \label{dif.cr.sec.}
\frac{d\sigma}{dt}={\frac{d\sigma}{dt}}^{Born}+\Phi(s,t,\lambda^2)\;,
\end{equation}
where the first term is the Born cross section and $\Phi(s,t,\lambda^2)$
takes into account the final-state interaction and its interference with
the Born part of the amplitude.

To establish the conditions under which
\begin{equation} \label{Phi-cond.}
\Phi(s,t,\lambda^2)=0\;,
\end{equation}
we use the existence theorem for implicit functions. By this theorem,
if equation (\ref{Phi-cond.}) allows the solution $s=s_0$, $t=t_0$,
$\lambda^2= \lambda^2_0$, and the function $\Phi(s,t,\lambda^2)$ and
its partial derivatives of the first order are continuous in the
vicinity of the point $M_0(s_0,t_0,\lambda^2_0)$ and the derivative
$\Phi_t^\prime$ at this point is different from zero, that is, 
\begin{equation} \label{Phi_t-cond.}
\Phi_t^\prime(s_0,t_0,\lambda^2_0)\neq 0\;,
\end{equation}
then there exists only one function
\begin{equation} \label{t=f} t=f(s,\lambda^2)\;,
\end{equation}
which satisfies equation (\ref{Phi-cond.}) in some vicinity of the
point $M_0$ and takes the value $t=t_0$ at $s=s_0$, $\lambda^2=\lambda^2_0$.
This function and its partial derivatives are continuous in the vicinity
of the point $M_0$.

According to this theorem, if we find at least one point in the space of
$(s,t,\lambda^2)$, where the effects of rescattering and their
interference with the Born terms compensate each other, then, since the
cross section is continuous in the physical region, there is a surface of
``compensation'' in this space on which the cross section is the Born one.
The intersections of this surface with every plane $\lambda^2=const$
define some curves in the plane $(s,t)$ each of them being characterized
by its own value of $\lambda^2$ and the cross section (\ref{dif.cr.sec.})
being the Born one along them. Thus, we obtain a one-parameter set of the
compensation curves with $\lambda^2$ as a parameter.

Consider the curves $\lambda^2=0$ in more detail. We shall use
$x=\cos\theta$ instead of variable $t$, where $\theta$ is a scattering
angle in the c.m. system. If the compensation takes place, then in the plane
$(x,s)$ we ought to have the compensation curve $x=f(s)$ continuous in the
physical region. Generally speaking, there may be several curves of this
kind, because the function $\Phi(s,x)$ can always be represented as a
polynomial in powers of $x$: $\Phi(s,x)=\sum_{i}^{}a_i(s)x^i~$ and the
equation $\sum_{i}^{}a_i(s)x^i=0$ can have several real roots
$x_j=f_j(s),~j=1,2,\cdots$. However, due to the theorem mentioned above, 
the curves do not intersect, since only one curve can pass through the
given point.

%
%
\begin{figure}[htb]
\begin{center}
\rotatebox{-90}{\scalebox{0.33}{\includegraphics{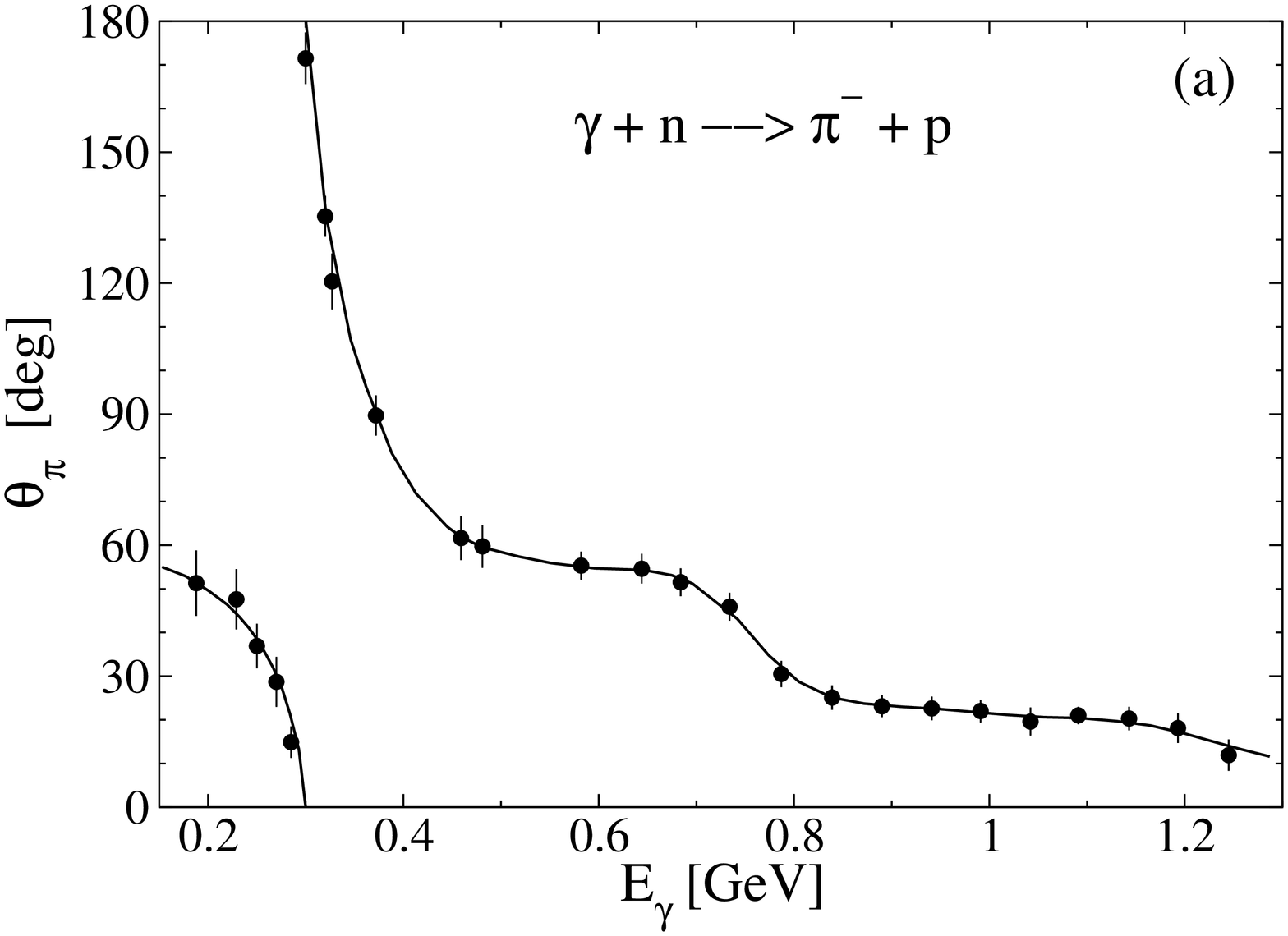}}}
\rotatebox{-90}{\scalebox{0.33}{\includegraphics{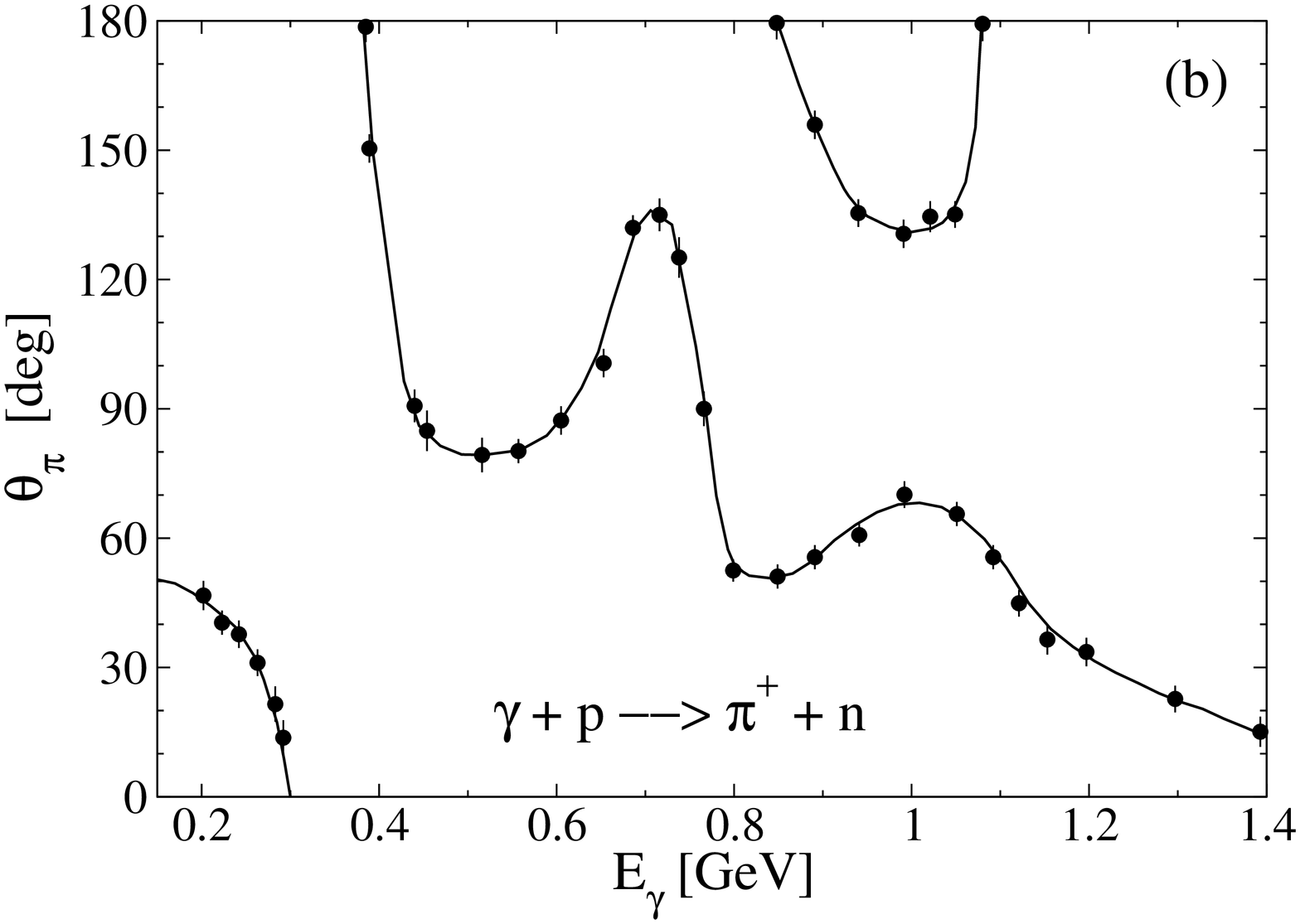}}}
\end{center}
\caption{Compensation curves for $\gamma p\to \pi^+ n$ and
$\gamma n\to \pi^- p$. Along these curves the differential cross section
is described by EBM. At high energies, the compensation curve passes
through the scattering angles corresponding to the value of the invariant
momentum transfer $t\approx -0.04({\rm GeV/c})^2$. }
\label{fig:comp-curv}
\end{figure}
In the case of the pion photoproduction, available experimental data are
sufficient to construct the compensation curves \cite{ST-yaf75}.
To this end, it is necessary to find out the intersection points of the
experimental differential cross section with the calculated Born cross
section and plot these points in the plane $(x,s)$ or $(\theta,s)$. In
Fig.~\ref{fig:comp-curv} the compensation curves for the charged pion
photoproduction are shown. Calculating the Born cross sections, we take
into account the interaction of the photon only with the charge.
Therefore, along these curves the experimental cross sections are
completely described by the electric Born terms. One can see that at high
energies, the compensation curves pass through small angles in accordance
with the electric Born model (EBM) \cite{Achasov,Cho-Sakurai}. The
compensation curves show also that an applicability of EBM to the cross
section extends to the whole energy range, but at the corresponding
angles, giving a more complete and transparent picture. Of course,
there are also compensation curves corresponding to the total Born
amplitude.

Now, if one assumes that for processes of the pion electroproduction and
IPE, when $\lambda^2\neq 0$, the compensation curves are not much
different from the corresponding curves for the photoproduction
$(\lambda^2=0)$, then along these curves the model dependence of
description of the processes should be minimal. In the resonance region,
this assumption would be reasonable if the change of $\lambda^2$ does not
lead to the essential reconstruction of the helicity (or multipole)
structure of the excitation of the corresponding resonance. In the first
resonance region, calculations, {\it e.g.}, in the dispersion theory,
which agree with the predictions of the quark models ({\it e.g.}
\cite{Goghilidze,Capstick,Close-Gilmal}) and with the result of the
phenomenological analysis of the electroproduction data ({\it e.g.}
\cite{Dev-Lyth}), confirm a dominance preservation of the magnetic dipole
excitation of the $P_{33}(1232)$ resonance with the change of
$\lambda^2 < 0$. Quark models expand this situation to values of
$\lambda^2 >0$. In our dispersion model, it was verified that the
compensation angle $\theta^{\gamma}_{comp}$ is invariable with $\lambda^2$
for the cross sections of the processes $\gamma^*n\leftrightarrow \pi^-p$
with the transverse virtual photons at $w =1.296$ GeV
$(\theta^{\gamma}_{comp} \simeq 70^0)$. This prediction was confirmed
by the experiment \cite{comp-exp} for the timelike values of
$\lambda^2$: $\theta^{\gamma}_{comp} = 70^0 \pm 7^0$. In the second and
third resonance regions, the analysis of experimental data for
electroproduction ({\it e.g.} \cite{Dev-Lyth}) gives a considerable change
(with $\vert\lambda^2\vert$) of the helicity structure for the excitation
of leading resonances, which is especially essential approximately at
$\lambda^2\simeq -1~({\rm GeV/c})^2$.

The compensation curves exist also for the asymmetry in the charged pion
production by the polarized $\gamma$-quanta
\begin{equation} \label{ph-asym}
\Sigma = [d{\sigma}_{\perp}/dt - d{\sigma}_{\parallel}/dt]/
[d{\sigma}_{\perp}/dt + d{\sigma}_{\parallel}/dt]\;.
\end{equation}
For example, at $w=1.296$~GeV the compensation in $\Sigma$ (\ref{ph-asym})
takes place at $\theta^{\gamma}_{comp} \simeq 30^0$ for the $\pi^{\pm}$
photoproduction.

Since at high energies and small $\vert t\vert$ EBM (as any reasonable
model explaining the sharp forward peak in $d\sigma /dt$ for the
$\pi^{\pm}$ photoproduction) gives the sign and size of
$\Sigma(\pi^{\pm})$ in the forward direction, the compensation curves for
$\Sigma(\pi^{\pm})$ above the resonance region go through approximately
the same values of the angles $(t\simeq 0.04 ({\rm GeV/c})^2)$ as for
$d\sigma(\pi^{\pm})/dt$. This fact is explained practically
model independently.

The plausibility of assumption that the compensation curves at high
energies and small angles and at $\lambda^2\neq 0$ are not much different
from the corresponding curves for $\lambda^2=0$ is justified by successful
application of EBM to the description of the processes of
electroproduction and $\pi N\to\rho^0 N$ at high energies and
$|t|\ltsim 2m_{\pi}^2$ \cite{Achasov,Cho-Sakurai}. In any case, the
compensation curves help one to reveal the optimal experimental conditions
for studying the form factors $F_{\pi}(\lambda^2)$ and $F_1^p(\lambda^2)$
in processes of the pion electroproduction and IPE.


\end{document}